\documentclass[fleqn,usenatbib]{mnras}
\usepackage{newtxtext,newtxmath}
\usepackage[T1]{fontenc}
\usepackage{ae,aecompl}
\usepackage{makecell}
\usepackage{graphicx}	
\usepackage{amsmath}	
\usepackage{amssymb}	
\usepackage{soul}
\usepackage{tabularx}
\usepackage{hyperref}

\newcommand{\Angstrom}{\textup{\AA}}
\newcommand{\skeltonphotz}{$z_{\mathrm{phot}}\texttt{(S14\_EAZY)}$}
\newcommand{\fixphotz}{$z_{\mathrm{phot}}\texttt{(R15fix)}$}
\newcommand{\rafelskiphotz}{$z_{\mathrm{phot}}\texttt{(R15\_BPZ)}$}
\newcommand{\bpz}{\textsc{bpz}}
\newcommand{\eazy}{\textsc{eazy}}
\newcommand{\betas}{$\langle\beta\rangle$}

\defcitealias{Schrabback2010EvidenceCOSMOS}{S10}
\defcitealias{Rafelski2015UVUDF:FIELD}{R15}
\defcitealias{Skelton20143D-HSTMASSES}{S14}
\defcitealias{Schrabback2018ClusterSurvey}{S18}

\title[Photometric redshift calibration for weak lensing]{Testing the accuracy of 3D-HST photometric redshift estimates as reference samples for deep weak lensing studies}

\author[S. F. Raihan et al.]{
S.~F.~Raihan,$^{1}$\thanks{E-mail: fraihan@astro.uni-bonn.de (KTS)}
T.~Schrabback,$^{1}$
H.~Hildebrandt,$^{2,1}$
D.~Applegate$^{1,3}$
and G.~Mahler$^{4}$
\\
$^{1}$Argelander-Institut f\"{u}r Astronomie, Universit\"{a}t Bonn, Auf dem
H\"{u}gel 71, 53121, Bonn, Germany\\
$^{2}$Astronomisches Institut, Ruhr-Universit\"{a}t Bochum, Universit\"{a}tsstr. 150, 44801, Bochum, Germany,\\
$^{3}$Kavli Institute for Cosmological Physics, University of Chicago, Chicago, IL 60637, USA,\\
$^{4}$Department of Astronomy, University of Michigan, 1085 South University Ave, Ann Arbor, MI 48109, USA
}

\date{Accepted 2020 June 30. Received 2020 June 25; in original form 2019 November 25}

\pubyear{2020}

\begin{document}

\label{firstpage}
\pagerange{\pageref{firstpage}--\pageref{lastpage}}
\maketitle

\begin{abstract}
Accurate weak lensing mass estimates of clusters are needed in order to calibrate mass proxies for the cosmological exploitation of galaxy cluster surveys. Such measurements require accurate knowledge of the redshift distribution of the weak lensing source galaxies. In this context, we investigate the accuracy of photometric redshifts (photo-$z$s) computed by the 3D-HST team for the Cosmic Assembly Near-infrared Deep Extragalactic Legacy Survey fields, which provide a relevant photometric reference data set for deep weak lensing studies. Through the comparison to spectroscopic redshifts and photo-$z$s based on very deep data from the
Hubble Ultra Deep Field, we identify catastrophic redshift outliers in the 3D-HST/CANDELS catalogue. These would significantly bias weak lensing results if not accounted for. We investigate the cause of these outliers and demonstrate that the interpolation of spectral energy distribution (SED) templates and a well-selected combination of photometric data can reduce the net impact for weak lensing studies.

\end{abstract}


\section{Introduction}

 Weak gravitational lensing is widely known as a tool to study the large-scale structure of the Universe. Distortions of light caused by massive objects within the cosmic web can be used to obtain unbiased estimates of the mass of these objects. One approach to constrain cosmology via weak lensing measurements is provided by cosmic shear \citep[e.g.][]{Schrabback2010EvidenceCOSMOS, Jee2016COSMICTOMOGRAPHY,Hildebrandt2018KiDS+VIKING-450:Data, Troxel2018SurveyMeasurements,vanUitert2018KiDS+GAMA:Clustering,Hikage2019CosmologyData, Chang2019ASurveys}. Another route where weak lensing measurements aid cosmological investigations are cluster number counts experiments \citep[e.g.][]{Reiprich2002TheClusters, Allen2011CosmologicalClusters}. In order to infer cosmological constraints from cluster surveys \citep[e.g.][]{Rozo2010COSMOLOGICALCATALOG, Mantz2014CosmologyConstraints, Schellenberger2017HICOSMO:Results, Bocquet2019ClusterTelescope/i}, we need to calibrate mass-observable scaling relations using weak lensing measurements over a broad redshift range. Galaxy cluster weak lensing mass calibration has so far mostly been done using ground-based data for low redshift clusters \citep[e.g.][]{Marrone2012LoCuSS:RELATION, vonderLinden2014RobustClusters, Hoekstra2015TheMasses, Applegate2016CosmologyLensing, Okabe2016LoCuSS:Clusters,  Stern2019Weak-lensingData, Dietrich2019SunyaevZeldovichClusters, McClintock2019DarkClusters} and space-based data studies of high redshift clusters \citep[e.g.][]{Leauthaud2010ARELATION, Hoekstra2011THECLUSTERS, Jee2011SCALINGTELESCOPE/i, Schrabback2018ClusterSurvey}. 
 
The South Pole Telescope Sunyaev-Zel'dovich (SPT-SZ) survey \citep{Bleem2015GALAXYSURVEY} has detected a large of sample massive clusters via their Sunyaev-Zel'dovich \citep[SZ;][]{Sunyaev1970TheSignificance,Sunyaev1972TheGalaxies} signal that extend out to high redshifts. \citet[S18 hereafter]{Schrabback2018ClusterSurvey} have performed a weak lensing analysis of a total of 13 high-redshift ($z_{\mathrm{median}}=0.88$) SPT-SZ galaxy clusters in order to aid the SPT-SZ cluster cosmology analysis \citep{Bocquet2019ClusterTelescope/i}.

To achieve accurate mass measurements from weak lensing, rigorous measurements of the shapes and redshift distributions of the weak lensing source galaxies must be obtained. The strength of weak lensing signals scales with the redshift-dependent geometric lensing efficiency. In order to accurately interpret the lensing signal and constrain mass models, accurate estimates of the source redshift distribution are therefore indispensable. Due to incompleteness at faint magnitudes, spectroscopic redshifts (spec-$z$s) of galaxies are typically insufficient to fully describe the redshift distribution for faint source samples. If there is sufficient photometric data, photometric redshifts (photo-$z$s) can be estimated directly from the weak lensing survey \citep[e.g.][]{deJong2013TheSurvey, Hildebrandt2017KiDS-450:Lensing}. To probe the weak lensing signal of high-redshift lenses, deeper data are needed, making it expensive to obtain observations in many bands for a large number of targets. A more cost-effective strategy is to obtain cluster field imaging in a few photometric bands only, which are chosen depending on the cluster redshift to facilitate an efficient selection of the main background source population via colour-cuts. A consistent colour selection must then be applied to photometric data from well-studied reference fields that had been covered over a wide range in wavelength, thus providing sufficient amount of photometric data to achieve reliable photo-$z$s \citep{Benitez2009OptimalEstimation}. In both cases a careful calibration of the inferred redshift distribution is required. This may employ deep spec-$z$s \citep[e.g.][]{Hildebrandt2018KiDS+VIKING-450:Data} and potentially higher quality photo-$z$s \citep[e.g.][]{Tanaka2018Photometric1}.

\citetalias{Schrabback2018ClusterSurvey} employed data from the Cosmic Assembly Near-infrared Deep Extragalactic Legacy Survey \citep[CANDELS;][]{Koekemoer2011CANDELS:MOSAICS, Grogin2011CANDELS:SURVEY, Galametz2013CANDELSFIELD} in order to calibrate the redshift distribution. Importantly, CANDELS includes deep $Hubble\ Space\ Telescope$ ($HST$) near-infrared (NIR) imaging, which greatly improves the photometric redshift estimation of $z\sim2$ galaxies by probing their $4000\Angstrom$/Balmer break. CANDELS also provides sufficient sky coverage over five lines of sight to suppress the impact of line-of-sight variations. Most of the CANDELS fields are also covered by the $HST$ $F814W$ and $F606W$ bands as needed for the colour selection that is used by \citetalias{Schrabback2018ClusterSurvey} to remove cluster galaxies and reduce foreground contamination.

In particular \citetalias{Schrabback2018ClusterSurvey} employed CANDELS photo-$z$s from 3D-HST \cite[S14 hereafter]{Skelton20143D-HSTMASSES} as reference sample. \citetalias{Schrabback2018ClusterSurvey} investigated the photo-$z$ accuracy through comparisons with the \textit{Hubble} Ultra Deep Field \citep[HUDF;][]{Beckwith2006TheField} data. They found significant issues with redshift outliers, for which they introduce an approximate empirical re-calibration scheme. The fact that this correction leads to a large ($12\%$) mass bias correction motivates further in-depth analysis of this important systematic issue, which is presented in this paper.

In Section \ref{section:photodata} we describe the data and catalogues that are used for our calibrations and tests. In Section \ref{section:S18}, we give a summary of the work done by \citetalias{Schrabback2018ClusterSurvey}, and describe the galaxy selection criteria that are relevant for this work. Section \ref{section:methods} details the comparison metric and the employed photo-$z$ codes. In Section \ref{section:results}, we present our main analysis and results, based on the comparison of different redshift samples, including various sets of photo-$z$s that are computed based on the 3D-HST photometric data with varying inputs and analysis schemes. In Section \ref{section:accuracy} we discuss the accuracy of our resulting redshift calibration and also compare it to the work done by \citetalias{Schrabback2018ClusterSurvey}. We summarise these findings in Section \ref{section:conclude}.

Throughout this paper we assume a standard flat $\Lambda$CDM cosmology characterised by $\Omega_\mathrm{m}=0.3$, $\Omega_\Lambda=0.7$, and $H_0=70h_{70}$ km/s/Mpc with $h_{70}=1$, as approximately consistent with CMB constraints \citep{Hinshaw2013NINE-YEARRESULTS}. All magnitudes are in the AB system and are corrected for galactic extinction according to \citet{Schlegel1998MapsForegrounds}.

\section{Photometric Data and Redshift Catalogues}
\label{section:photodata}

\subsection{3D-HST}
\label{subsection:3dhst}

3D-HST \citep{Brammer20123D-HST:TELESCOPE/i} is a 248-orbit $HST$ treasury programme that builds upon the CANDELS programme by adding $HST$ Wide Field Camera 3 (WFC3) G141 grism observations for slitless spectroscopy across $75\%$ of the CANDELS area. Apart from the grism slitless spectroscopy, 3D-HST also yields WFC3 $F140W$ and Advanced Camera for Surveys(ACS) $F814W$ imaging data in parallel. In \citetalias{Skelton20143D-HSTMASSES}, the 3D-HST team presents photo-$z$s for the five CANDELS fields, employing deep photometric data from $HST$ and ancillary imaging data, with at least five $HST$ photometric bands in each field. They also released their grism slitless spectroscopic redshift estimates (grism-$z$s) and a compilation of spec-$z$s from ground-based programmes \citep{Momcheva2016THEGALAXIES}. \citetalias{Skelton20143D-HSTMASSES} calculated their photo-$z$s using \eazy ~\citep{Brammer2008EAZY:Code}, which is an algorithm that estimates photo-$z$ using the spectral energy distribution (SED) template-fitting technique. 
The SED templates correspond to the default set described in \cite{Brammer2008EAZY:Code}, which contains four templates derived from a library of P\'EGASE stellar population models \citep{Fioc1997PEGASE:Counts}, a young, dusty galaxy template and a red galaxy template, as described in \citet{Whitaker2011THE3}. The red galaxy template is derived from the \citet{Maraston2005EvolutionaryGalaxies} stellar population synthesis models with an age of $12.6$Gyr, a Kroupa initial mass function (IMF) and solar metallicity. \citetalias{Skelton20143D-HSTMASSES} also modify the templates in the fitting procedure by correcting for subtle differences between the observed SEDs of galaxies and the best-fitting templates. The redshift prior used by \citetalias{Skelton20143D-HSTMASSES} is based on the $K$-band apparent magnitude coming from the $K$-band number counts in the light-cone simulation of \citet{Blaizot2005MoMaF:Facility}.

\subsection{UVUDF}
\label{subsection:UVUDF}

The UVUDF \citep{Teplitz2013UVUDF:3} is an $HST$ programme (GO-12534; PI: Teplitz) that obtained deep, near-ultraviolet (NUV) imaging of the HUDF. The HUDF benefit from large spectral coverage, reaching $28.3$ mag depth for the NUV bands and $29.8$ mag for the optical/NIR. The NUV $F225W$, $F275W$ and $F336W$ bands improve the redshift estimates by sampling the Lyman break of high-redshift galaxies and the Balmer or $4000\Angstrom$ break for low-redshift galaxies. The optical data is provided by the four original ACS $F435W$, $F606W$, $F775W$, and $F850LP$ filters \citep{Beckwith2006TheField}. \citet[R15 hereafter]{Rafelski2015UVUDF:FIELD} have released the photo-$z$s calculated using the acquired NUV data as well as NIR data from the UDF09 and UDF12 programmes \citep{Oesch2010StructureField,Oesch2010iz/iRESULTS, Bouwens2011UltravioletObservations,Ellis2013THECAMPAIGN,Koekemoer2013THEOVERVIEW} and the CANDELS GOODS-S programmes \citep{Grogin2011CANDELS:SURVEY, Koekemoer2011CANDELS:MOSAICS}, building upon the photo-$z$s presented in \cite{Coe2006GalaxiesMorphology}. \citetalias{Rafelski2015UVUDF:FIELD} also compiled matching spec-$z$s from various ground-based programmes, when available. 
We exploit the greater depth and the greater wavelength range coverage of the UVUDF to provide us with another avenue to test and calibrate photo-$z$s, particularly from \citetalias{Skelton20143D-HSTMASSES}. This is due to the spatial overlap of the HUDF and the CANDELS/GOODS-S field. These deeper photo-$z$s provide a reference sample that does not suffer from the incompleteness issues of spectroscopic and grism samples.

\citetalias{Rafelski2015UVUDF:FIELD} calculated photo-$z$s using two SED template fitting codes, \bpz ~\citep{Benitez2000BayesianEstimation} and \eazy. After comparing, they found that the photo-$z$s computed using \bpz ~perform slightly better in terms of scatter and outlier fraction than the ones from \eazy. The \bpz ~SED templates used by \citetalias{Rafelski2015UVUDF:FIELD} are described in \citet{Coe2006GalaxiesMorphology}. The SED set consists of four elliptical galaxies (Ell), one Lenticular (ESO), two spirals (Sbc and Scd), and four starbursts (SB). These templates are based on those from P\'EGASE \citep{Fioc1997PEGASE:Counts} but re-calibrated based on observed photometry and spec-$z$s from FIREWORKS \citep{Wuyts2008FIREWORKSGalaxies}. \citetalias{Rafelski2015UVUDF:FIELD} also created nine intermediate templates that were interpolated between adjacent templates through the "\texttt{INTERP}" function in \bpz. Overall \citetalias{Rafelski2015UVUDF:FIELD} employ 111 galaxy templates. The redshift prior used by \citetalias{Rafelski2015UVUDF:FIELD} is the default prior in \bpz ~which is based on the $F814W$-band apparent magnitude calibrated using HDF-N \citep{Williams1996ThePhotometry} and CFRS \citep{Lilly1995TheEffects, Crampton1995THESample} spec-$z$ catalogues. 

\subsection{MUSE Hubble Ultra Deep Field Survey}

The Multi-Unit Spectroscopic Explorer \citep[MUSE;][]{Bacon2010TheInstrument} is an integral field spectrograph (IFS) on the Very Large Telescope (VLT). The MUSE Hubble Ultra Deep Field Survey \citep{Bacon2015TheSouth} presented spec-$z$s from MUSE in the HUDF in \cite{Inami2017TheSurvey}. The galaxies from \cite{Inami2017TheSurvey} were matched to galaxies detected in the \citetalias{Rafelski2015UVUDF:FIELD} catalogue and a detailed photo-$z$s calibration to the 30\textsuperscript{th} magnitude has been presented in \citet{Brinchmann2017Survey}. MUSE has boosted the number of spec-$z$ of detected galaxies in the HUDF from $2\%$ quoted in \citetalias{Rafelski2015UVUDF:FIELD} to $15\%$. Among our 261 colour and magnitude-selected galaxies (see Section \ref{section:S18}), $49\%$ now have spectroscopic or grism redshifts ($z_{\mathrm{spec/grism}}$).

\subsection{Matching the catalogues}

The catalogues are matched based on their coordinates using the function \verb|associate| that is part of the \verb|LDAC|\footnote{\url{http://marvinweb.astro.uni-bonn.de/data_products/THELIWWW/LDAC}} tools, requiring that the catalogue positions differ by less than $0\farcs1$\footnote{We have verified that larger matching radii do not provide significant advantage. E.g., using $0\farcs3$ matching radius increases the number of matched galaxies by $\sim1\%$ only.}. We then extract the matched galaxies with their corresponding photometric data from the \citetalias{Skelton20143D-HSTMASSES} catalogue\footnote{\url{https://3dhst.research.yale.edu/Data.php}} and the \citetalias{Rafelski2015UVUDF:FIELD} catalogue\footnote{\url{http://uvudf.ipac.caltech.edu/catalogs.html}}. MUSE spec-$z$s from \cite{Inami2017TheSurvey} are then extracted using the \citetalias{Rafelski2015UVUDF:FIELD} identification number in the catalogue\footnote{\url{http://cdsarc.u-strasbg.fr/viz-bin/qcat?J/A+A/608/A2}}.

\section{Summary of the S18 Redshift Calibration}
\label{section:S18}

In this section we summarise the work done in \citetalias{Schrabback2018ClusterSurvey} regarding redshift distribution calibration, as this marks the starting point for our investigation.

The aim of this study is to calibrate the redshift data from \citetalias{Skelton20143D-HSTMASSES}, so that they can be used to accurately estimate the source redshift distribution for weak lensing studies such as the one conducted by \citetalias{Schrabback2018ClusterSurvey}. The lenses studied in \citetalias{Schrabback2018ClusterSurvey} are high-redshift galaxy clusters (\mbox{$0.6 \lesssim z \lesssim 1.1$}). At the cluster redshifts, the field is over-dense and does not represent the cosmic mean distribution of galaxies. Therefore, \citetalias{Schrabback2018ClusterSurvey} applied cuts based on magnitude \mbox{$24<V_{\mathrm{606}}<26.5$} and colour \mbox{$V_{\mathrm{606}}-I_{\mathrm{814}}<0.3$}\footnote{\citetalias{Schrabback2018ClusterSurvey} use this colour cut for clusters with redshifts \mbox{$z_{\mathrm{c}}<1.01$}, and \mbox{$V_{\mathrm{606}}-I_{\mathrm{814}}<0.2$} at \mbox{$z_{\mathrm{c}}>1.01$}. They also employ slightly bluer cuts for noisy galaxies to keep cluster contamination low.} to remove both red and blue galaxies in the redshift range of the galaxy clusters. The study by \citet{Tholken2018iXMM-Newton/i0.902} has also used this colour-cut method. It is important to mimic these selection criteria in the reference fields that are used to estimate the source redshift distribution. 

Through comparison with $z_{\mathrm{spec/grism}}$, \citetalias{Schrabback2018ClusterSurvey} found that \citetalias{Skelton20143D-HSTMASSES} photo-$z$s are reasonably well calibrated but suffer from a few systematic features. First, the comparison to the $z_{\mathrm{spec/grism}}$ revealed the presence of catastrophic redshift outliers, mostly in the form of galaxies at \mbox{$2<z_{\mathrm{spec/grism}}<3$} that are assigned a redshift below $0.3$ (see also Figure \ref{fig:1} and also Section \ref{section:results}). This might be due to inaccurate template matching and degeneracies of the colour-redshift relation. Other issues that are not as prominent are noticeable redshift focusing \citep{Wolf2009BayesianSets} at \mbox{$1.4<$ photo-$z<1.6$} and the observation that the \citetalias{Skelton20143D-HSTMASSES} photo-$z$s tend to generally be biased low compared to $z_{\mathrm{spec/grism}}$ at \mbox{$z_{\mathrm{spec/grism}}>2.5$}.

\citetalias{Schrabback2018ClusterSurvey} found that the \citetalias{Rafelski2015UVUDF:FIELD} photo-$z$s, calculated using \bpz, performed better than the \citetalias{Skelton20143D-HSTMASSES} photo-$z$s in terms of catastrophic outliers and the overall distribution, although \citetalias{Rafelski2015UVUDF:FIELD} slightly overestimated photo-$z$s in the intervals of \mbox{$1.0<$ photo-$z<1.7$} and \mbox{$2.6<$ photo-$z<3.7$}. Upon fixing the overestimated photo-$z$s by shifting the \citetalias{Rafelski2015UVUDF:FIELD} photo-$z$s in these two intervals, \citetalias{Schrabback2018ClusterSurvey} found that the fixed \citetalias{Rafelski2015UVUDF:FIELD} photo-$z$s provide a sufficiently good approximation of the true redshifts. Hence, \citetalias{Schrabback2018ClusterSurvey} utilised the fixed \citetalias{Rafelski2015UVUDF:FIELD} photo-$z$s as a reference to obtain a statistical correction for the systematic features of the \citetalias{Skelton20143D-HSTMASSES} photo-$z$s.

\citetalias{Schrabback2018ClusterSurvey} had applied a statistical ad-hoc calibration method to correct for the systematic bias in the  photo-$z$s distribution. The procedure is as follows; for each \citetalias{Skelton20143D-HSTMASSES} galaxy with \mbox{photo-$z<0.3$} and \mbox{$V_{606}-I_{814}<0.2$} \citetalias{Schrabback2018ClusterSurvey} add a randomly drawn offset from the comparison between \citetalias{Skelton20143D-HSTMASSES} and the fixed \citetalias{Rafelski2015UVUDF:FIELD} redshifts to its photo-$z$s. \citetalias{Schrabback2018ClusterSurvey} also apply a statistical correction for the redshift focusing within the redshift range \mbox{$1.4<$ photo-$z<1.6$} for galaxies with \mbox{$V_{606}-I_{814}<0.1$}, which are seen as most strongly affected, again randomly sampling from the corresponding \mbox{$(z_\mathrm{R15,fix}-z_\mathrm{S14})_i$} offsets in the HUDF. For the latter correction \citetalias{Schrabback2018ClusterSurvey} split the galaxies into two magnitude ranges (\mbox{$24<V_{606}<25.5$} and \mbox{$25.5<V_{606}<26.5$}) given that the fainter galaxies appear to suffer from the redshift focusing effects more strongly. They have found that this statistical correction has resulted in a more unbiased distribution of photo-$z$s when compared against \citetalias{Rafelski2015UVUDF:FIELD}. The remaining systematic uncertainty of this correction estimated by \citetalias{Schrabback2018ClusterSurvey} is $2.2\%$ (in terms of the average geometric lensing efficiency) for the photo-$z$s with an additional $1\%$ to account for variation between CANDELS fields.

 In this paper, we investigate the origin of the outliers in the photo-$z$s calculated by \citetalias{Skelton20143D-HSTMASSES} with an emphasis on the source population in weak lensing studies of high-redshift galaxy clusters. We also provide a more robust solution to correct for the systematic bias. We will mainly show results for the colour-magnitude-selected sample used in the analysis in \citetalias{Schrabback2018ClusterSurvey}. We will also show results for a purely magnitude-selected galaxy sample with \mbox{$23<I_{\mathrm{814}}<27$} in Appendix \ref{app:magselect}, which gives us an added insight to our analysis and might be of relevance for other weak lensing studies such as cosmic shear.

\section{Methods}
\label{section:methods}
In this section, we explain the comparison metrics, as well as the error calculation that we use to evaluate the photo-$z$s quantitatively. Then, we elaborate on how we re-calculate the photo-$z$s using the algorithms \bpz ~and \eazy.

\subsection{Comparison metrics}
\label{subsection:comparison methods}

For our weak lensing analysis the most relevant metric of the redshift distribution is given by the mean geometric lensing efficiency, \betas, which is defined as, 
\begin{equation}
\langle\beta\rangle=\frac{\sum  \beta(z_i) w_i}{\sum w_i} \, ,\, 
\end{equation}
where,
\begin{equation}
\beta=\mathrm{max}\left[0,\frac{D_\mathrm{ls}}{D_\mathrm{s}}\right] .
\end{equation}
Here, $w$ is the magnitude-dependent shape weight that is obtained through empirical fitting by \citetalias{Schrabback2018ClusterSurvey}\footnote{S18 apply the same colour and magnitude selection to CANDELS galaxies and measure PSF-corrected weak lensing galaxy ellipticities $\epsilon$ from stacks of approximately single-orbit depth, matching the depth of the cluster field observations. Splitting the galaxies into magnitude bins they then fit the magnitude-dependent ellipticity dispersion $\sigma_\epsilon (V_{606})$, which yields the shape weights as $w_i=[\sigma_\epsilon (V_{606,i})]^{-2}$.} that down-weights contributions from faint galaxies, and $D_\mathrm{s}$ and $D_\mathrm{ls}$ indicate the angular diameter distances to the source and between the lens and the source, respectively. This is the most relevant metric of comparison for our studies because the cluster tangential shear scales with \betas. Throughout this study, we assume the lens redshift to be at \mbox{$z=0.9$} in accordance with the mean redshift of clusters in \citetalias{Schrabback2018ClusterSurvey}.

We then compare the \betas ~values computed from the investigated redshift distribution against a reference value, i.e. $\langle\beta_{\mathrm{ref}}\rangle$, computed from the reference distribution. We derive the relative bias of the mean geometric lensing efficiency by dividing the difference between the test \betas ~and $\langle\beta_{\mathrm{ref}}\rangle$ by $\langle\beta_{\mathrm{ref}}\rangle$. 
The errors on the relative bias are calculated by bootstrapping\footnote{We did not adopt the spatial bootstrapping approach that is employed by \citet[]{Hildebrandt2018KiDS+VIKING-450:Data}} our galaxy sample. For each bootstrap sample realisation, we compute the \betas ~and $\langle\beta_{\mathrm{ref}}\rangle$ before taking the difference and normalising over $\langle\beta_{\mathrm{ref}}\rangle$.

The size of the error also depends on the whole distribution of redshifts. Considering the same sample size, a larger error means that there is a bigger dispersion in the bootstrapped calculation of the relative bias, which indicates that there is a bigger dispersion of \betas$-\langle\beta_{\mathrm{ref}}\rangle$. This means that for a larger error, there is more overall scatter. For the cluster mass calibration, an error of $\lesssim2\%$ is to be desired.

Another metric of comparison for the photo-$z$s is the catastrophic outlier fraction (COLF), where we defined a galaxy as outlier if it obeys \mbox{$\left|\Delta z\right|/(1+z_{\mathrm{spec}})>0.6$}. A high COLF is a sign that the photo-$z$s may be problematic, albeit being sometimes less biased than photo-$z$s with low COLF.

\subsection{Photometric redshift algorithms}

To investigate the cause of the difference of the \citetalias{Skelton20143D-HSTMASSES} and \citetalias{Rafelski2015UVUDF:FIELD} photo-$z$s, we re-calculate the photo-$z$s by employing the two template fitting algorithms, \eazy ~\citep{Brammer2008EAZY:Code} which was applied by both \citetalias{Skelton20143D-HSTMASSES} and \citetalias{Rafelski2015UVUDF:FIELD}, and \bpz ~\citep{Benitez2000BayesianEstimation}, which was also used by \citetalias{Rafelski2015UVUDF:FIELD}. The photo-$z$s from both algorithms should match each other although some difference is possible. We summarise some of the most important differences between \eazy ~and \bpz ~below and refer to \citet{Hildebrandt2010PHAT:Testing} for further details.

 The \bpz ~and \eazy ~codes both implement template-fitting algorithms to fit photometric data to a set of redshifted SEDs of galaxy templates. By integrating the likelihood employing a Bayesian prior they then determine the resulting photo-$z$s. For more details on template-fitting methods, see e.g. \citet{Bolzonella2000PhotometricProcedures, Benitez2000BayesianEstimation, Hildebrandt2010PHAT:Testing}.

We employ the same \eazy ~version\footnote{We managed to get the exact version the \citetalias{Skelton20143D-HSTMASSES} team used through private communication.} that \citetalias{Skelton20143D-HSTMASSES} utilised. This includes all the SED templates, priors and settings discussed in detail in Section \ref{subsection:3dhst}.

We use the public \bpz ~package\footnote{\url{http://www.stsci.edu/~dcoe/BPZ/} version 1.99.3}, which employs an SED templates set that concentrates on the properties of high-redshift galaxies. It includes one elliptical, two spirals and an irregular type from \cite{Coleman1980COLORSGALAXIES}, two starburst galaxies from \cite{Kinney1996TemplateK-Corrections}, and two steep "blue" $25$Myr and $5$Myr simple stellar population model SEDs from \cite{Bruzual2003Stellar2003} that have been added to accommodate the large population of faint blue galaxies observed in the HUDF. An important feature in \bpz ~is the interpolation between galaxy templates to mitigate the problem of incomplete template sets. The redshift prior that is employed in \bpz ~is the default $I$-band based prior (see Section \ref{subsection:UVUDF}). 

\section{Results}
\label{section:results}

\begin{figure*}
 \includegraphics[width=0.68\columnwidth]{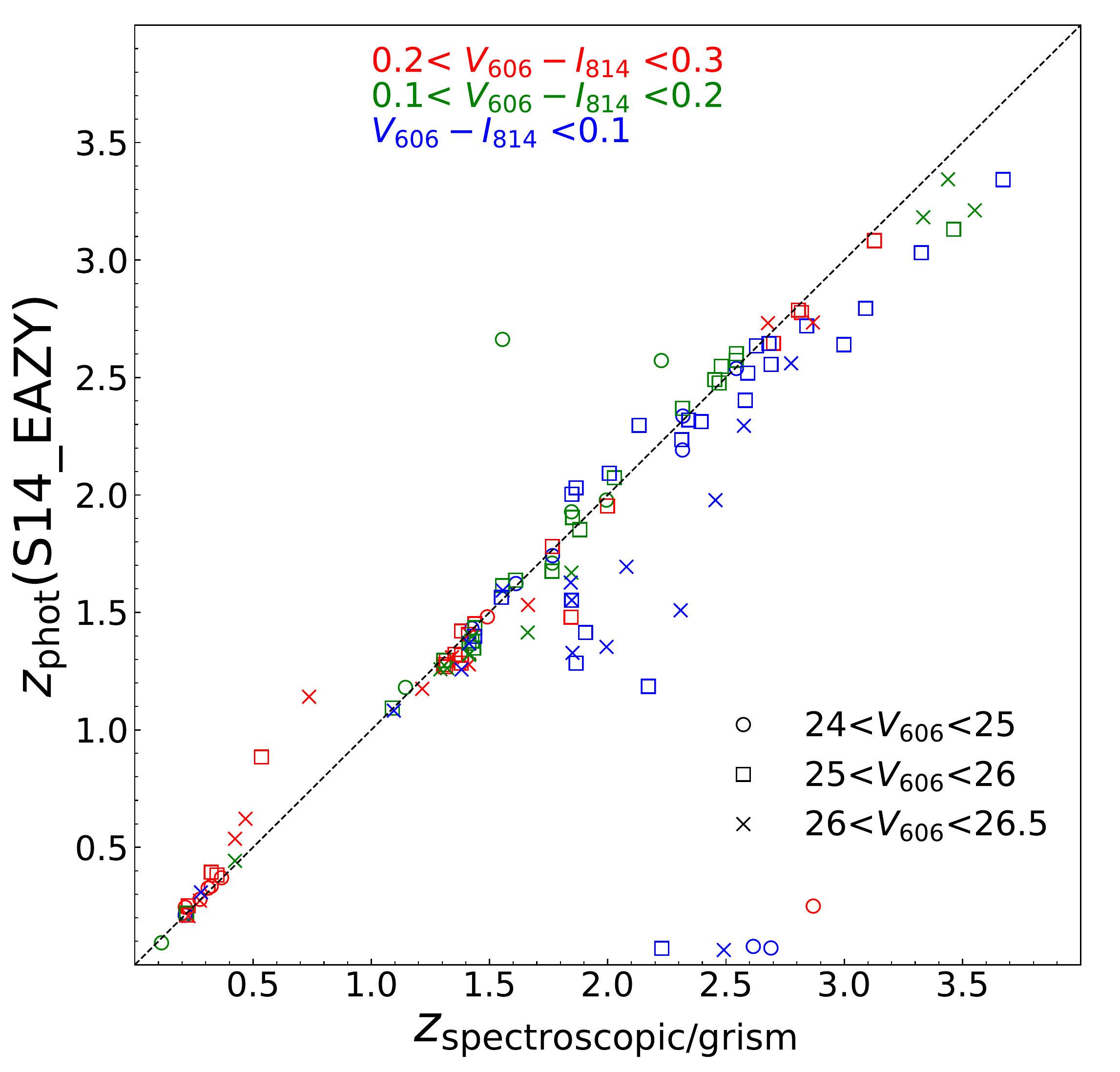}
 \includegraphics[width=0.68\columnwidth]{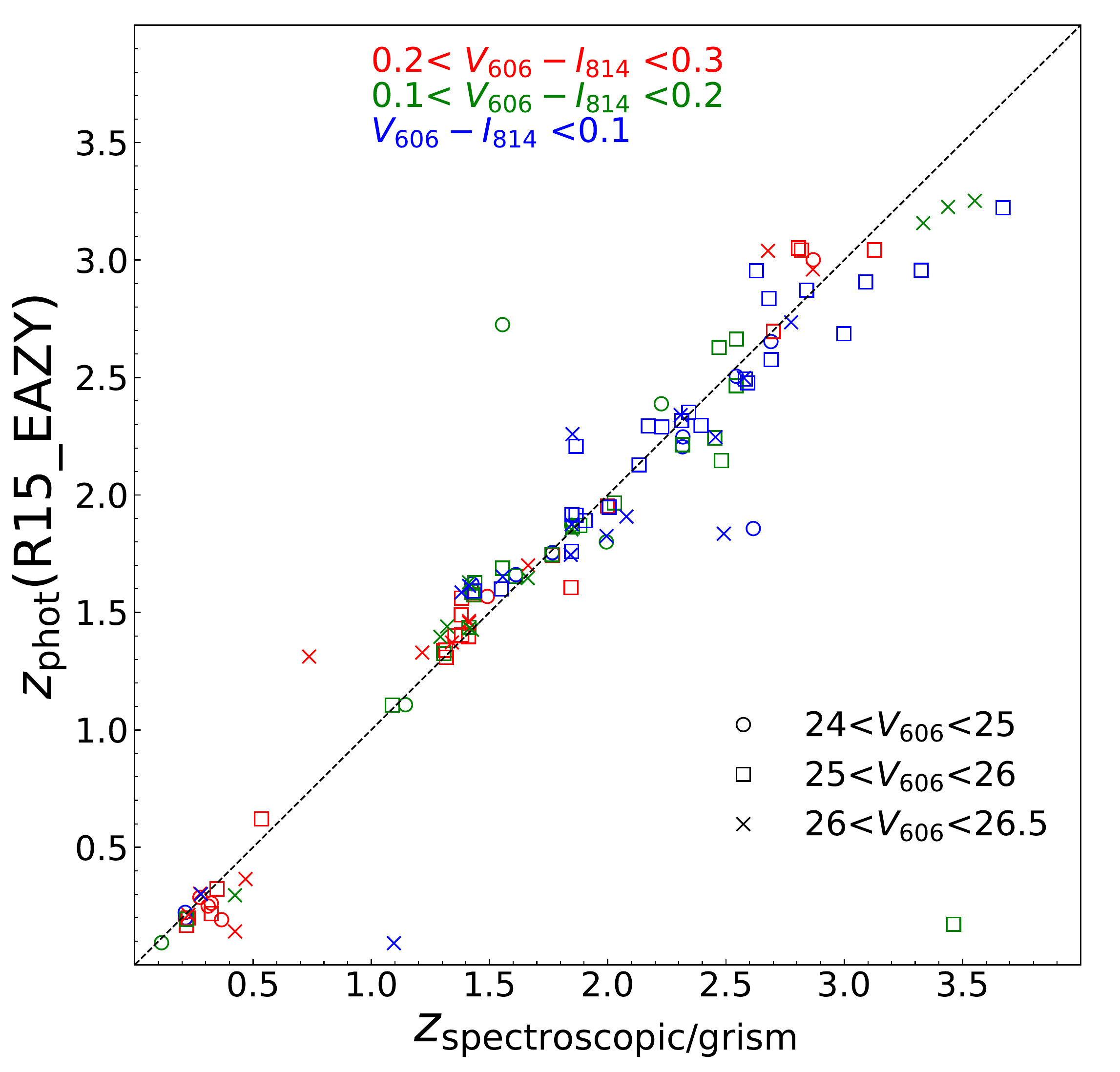}
  \includegraphics[width=0.68\columnwidth]{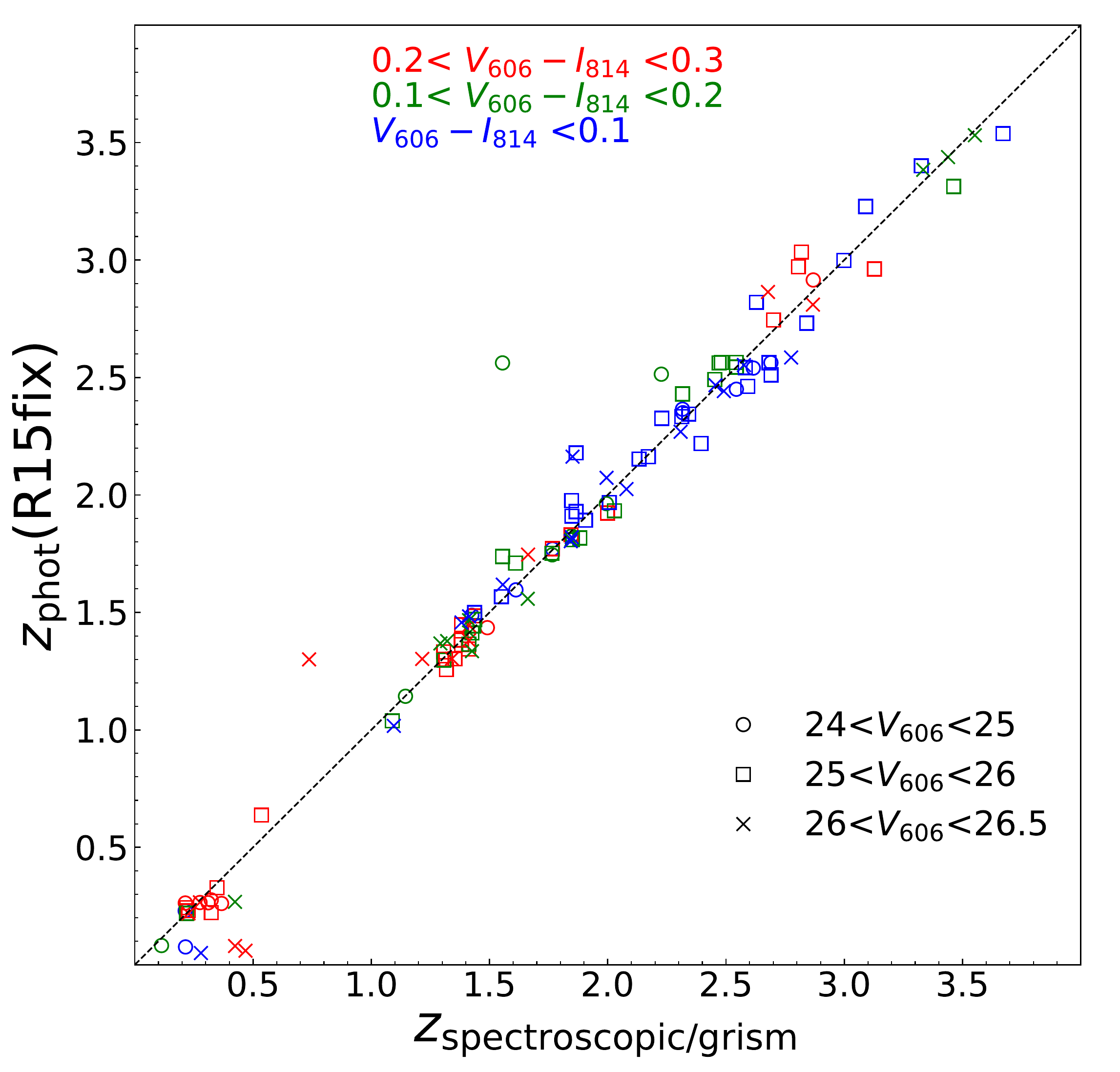}
\caption {Comparison of photo-$z$s in the HUDF including the peak photo-$z$s from \eazy ~computed by the 3D-HST team in the GOODS-South field ({\it left}), the \eazy ~photo-$z$s from  the UVUDF project ({\it middle}), the \bpz ~photo-$z$s from the UVUDF project ({\it right}, with small bias corrections applied, see text) to $z_{\mathrm{spec/grism}}$. Different colours and symbols correspond to different colour and magnitude ranges as indicated (based on the 3D-HST photometry).}
\label{fig:1}
\end{figure*}

\subsection{S14 and R15 photo-$z$s comparison revisit}
\label{subsection:S14R15compare}

In this section we revisit the \citetalias{Rafelski2015UVUDF:FIELD} and \citetalias{Skelton20143D-HSTMASSES} photo-$z$s comparison, now including redshifts from MUSE.

In the \citetalias{Schrabback2018ClusterSurvey} comparison, \citetalias{Schrabback2018ClusterSurvey} used the \cite{Momcheva2016THEGALAXIES} grism redshifts and \citetalias{Skelton20143D-HSTMASSES} spec-$z$s compilation that includes 66 redshifts for the colour-and-magnitude-selected sample. Through the addition of the MUSE spec-$z$s \citep{Inami2017TheSurvey} we have been able to include 128 galaxies in total. We perform the redshift comparison of the \eazy ~photo-$z$s from \citetalias{Skelton20143D-HSTMASSES} and \citetalias{Rafelski2015UVUDF:FIELD}, denoted as \skeltonphotz ~and $z_{\mathrm{phot}}\texttt{(R15\_EAZY)}$, to the $z_{\mathrm{spec/grism}}$ in Figure \ref{fig:1}. Similarly to \citetalias{Schrabback2018ClusterSurvey}, we find that the \skeltonphotz ~are reasonably well calibrated but still show the same systematic features as stated in Section \ref{section:S18}. We find that the \rafelskiphotz ~still perform better, but are slightly biased high compared to $z_{\mathrm{spec/grism}}$. The bias in the photo-$z$s ranges \mbox{$1.0<z<1.7$}(\mbox{$2.6<z<3.2$}) amounts to $0.081$($0.162$). We subtract the median offsets in these ranges ($0.081$ and $0.162$, respectively) from \rafelskiphotz, yielding \fixphotz, which is shown in Figure \ref{fig:1}. For comparison, $z_{\mathrm{phot}}\texttt{(R15\_EAZY)}$ also shows fewer catastrophic outliers than \skeltonphotz ~but suffers from a larger scatter around the one-to-one line compared to \rafelskiphotz.

\begin{figure*}
 \includegraphics[width=0.6\linewidth]{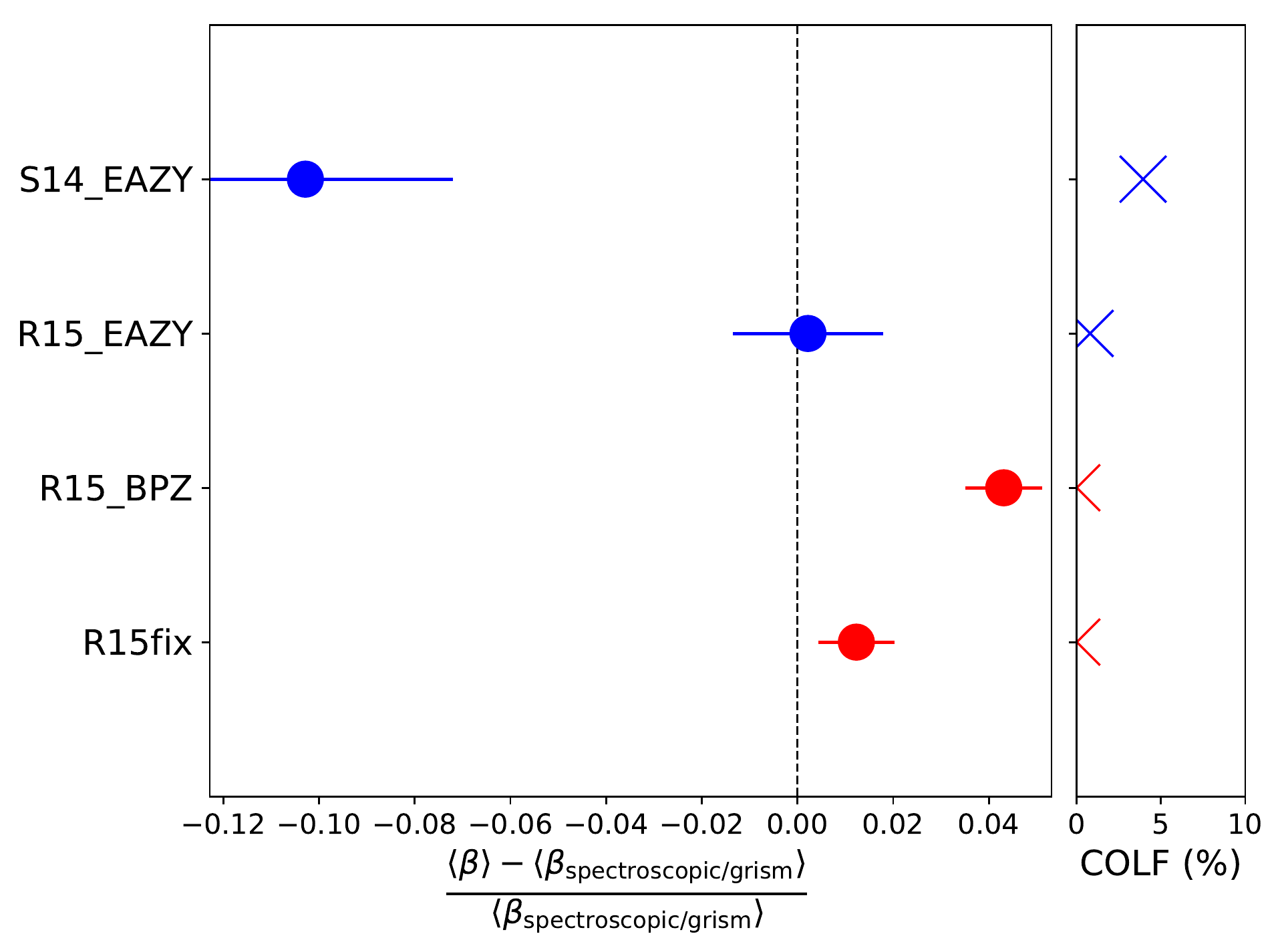}
\caption{The relative bias in the mean geometric lensing efficiency normalised to the $z_{\mathrm{spec/grism}}$ for the galaxies in our colour-and-magnitude-selected sample that have a match in the spec/grism-$z$s catalogue. The errors are determined via bootstrapping (see text). The right panel shows the corresponding COLF (see text).}
\label{fig:nmad1}
\end{figure*}

In Figure \ref{fig:nmad1} we show the relative bias in \betas ~of \skeltonphotz, $z_{\mathrm{phot}}\texttt{(R15\_EAZY)}$, \rafelskiphotz, and \fixphotz ~compared to $z_{\mathrm{spec/grism}}$ along with their corresponding COLF. We see that \skeltonphotz ~is indeed strongly biased low with a higher COLF compared to the rest. \rafelskiphotz ~is biased high but with zero COLF. $z_{\mathrm{phot}}\texttt{(R15\_EAZY)}$ has a low COLF and \betas ~consistent with unbiased, however looking at the size of the error bar, it is much bigger than the error of \rafelskiphotz, which indicates that it has more scatter, as evident in Figure \ref{fig:1}. To establish a new calibration sample, a tight correlation to $z_{\mathrm{spec/grism}}$ is desired. Therefore, we choose \fixphotz, which is only slightly biased high with a smaller error bar. This corresponds to a $\sim+1\%$ overestimation of the \betas, which we will discuss and take into account in Section \ref{subsection:simulateshallow}. Importantly,  \fixphotz ~does not suffer from the issue of catastrophic redshift outliers discussed in Section \ref{section:S18}, which is essential for its use as a robust reference sample. In contrast, a significant, non-zero catastrophic redshift outlier fraction could depend on selection effects, which might differ between spectroscopic and photometric galaxy samples and lead to biased redshift calibrations when derived from incomplete reference samples. 

However, we cannot exclude that spectroscopic selection effects might slightly affect our correction of the \rafelskiphotz ~to \fixphotz. For a conservative sensitivity analysis we assume a scenario in which the correction is preferentially required for galaxies with a spec/grism-$z$ estimate (on average these galaxies likely show stronger emission lines than random galaxies). If we conservatively assume that no correction would be required for half of the galaxies without  spec/grism-$z$s, \betas ~would shift by $0.8\%$ only. We include this as a systematic uncertainty estimate in our final systematic error budget in Section \ref{subsection:accountingfieldvariation}.

\begin{figure*}
\includegraphics[width=0.68\columnwidth]{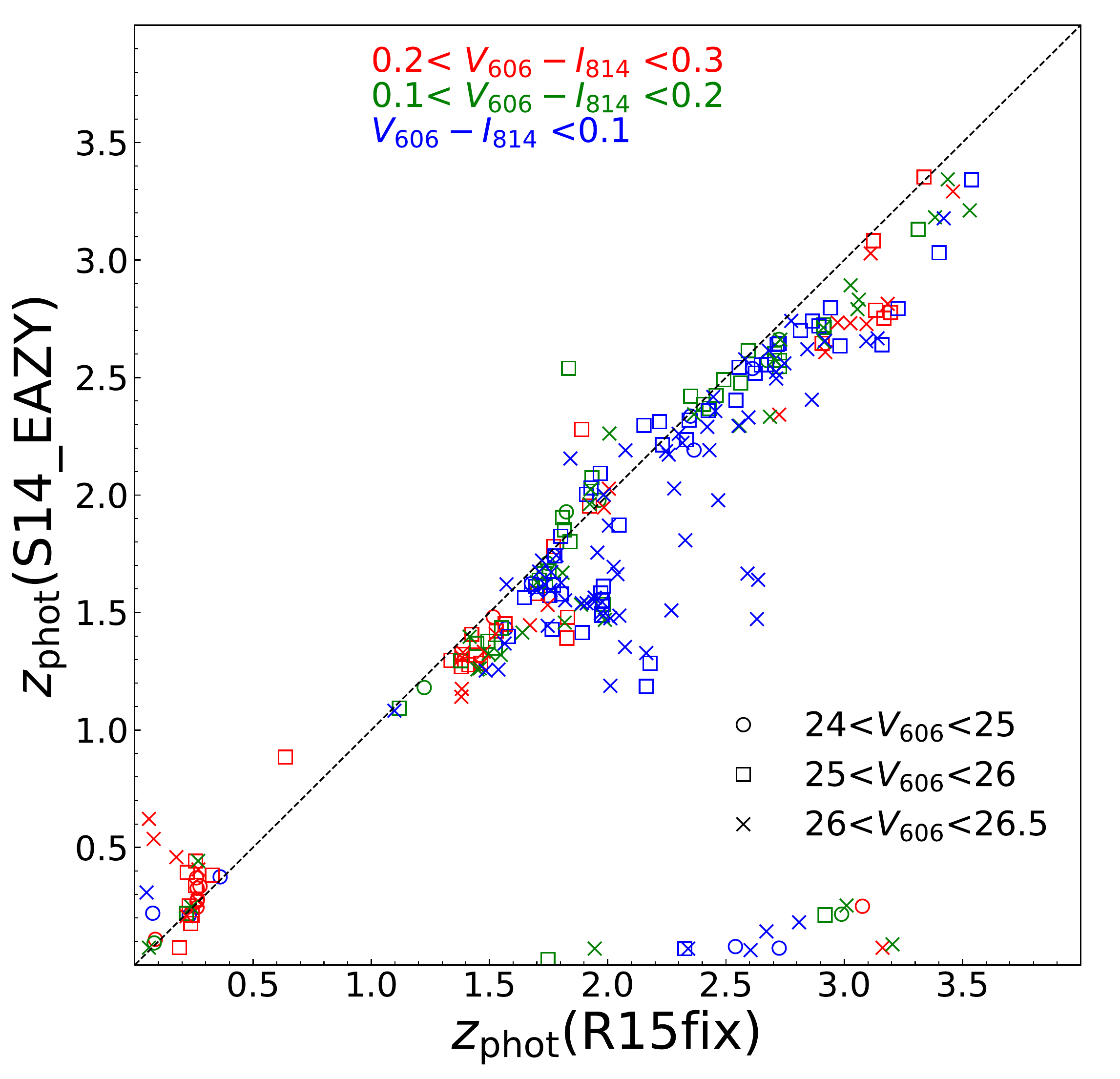}
\includegraphics[width=0.68\columnwidth]{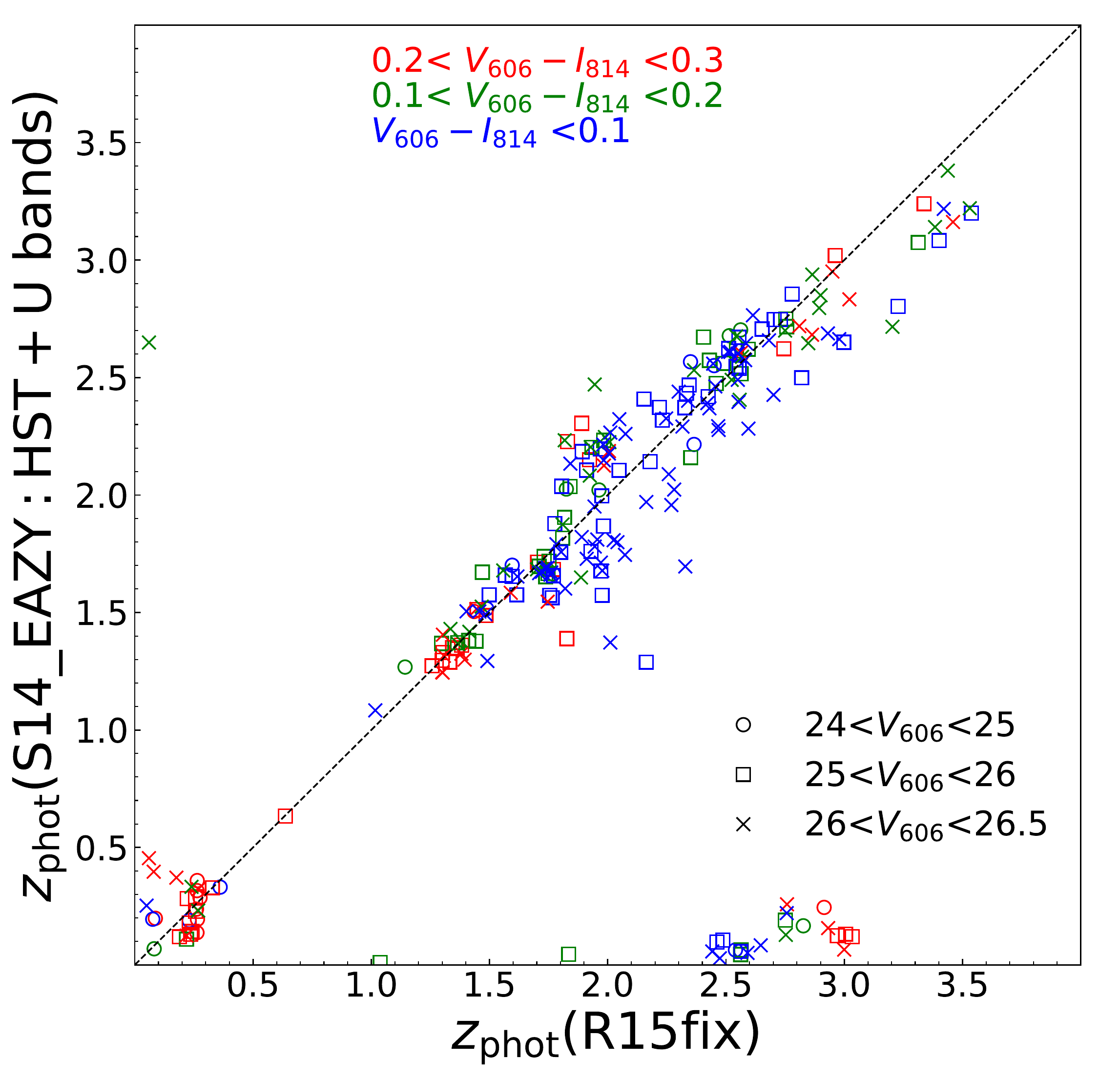}
\includegraphics[width=0.68\columnwidth]{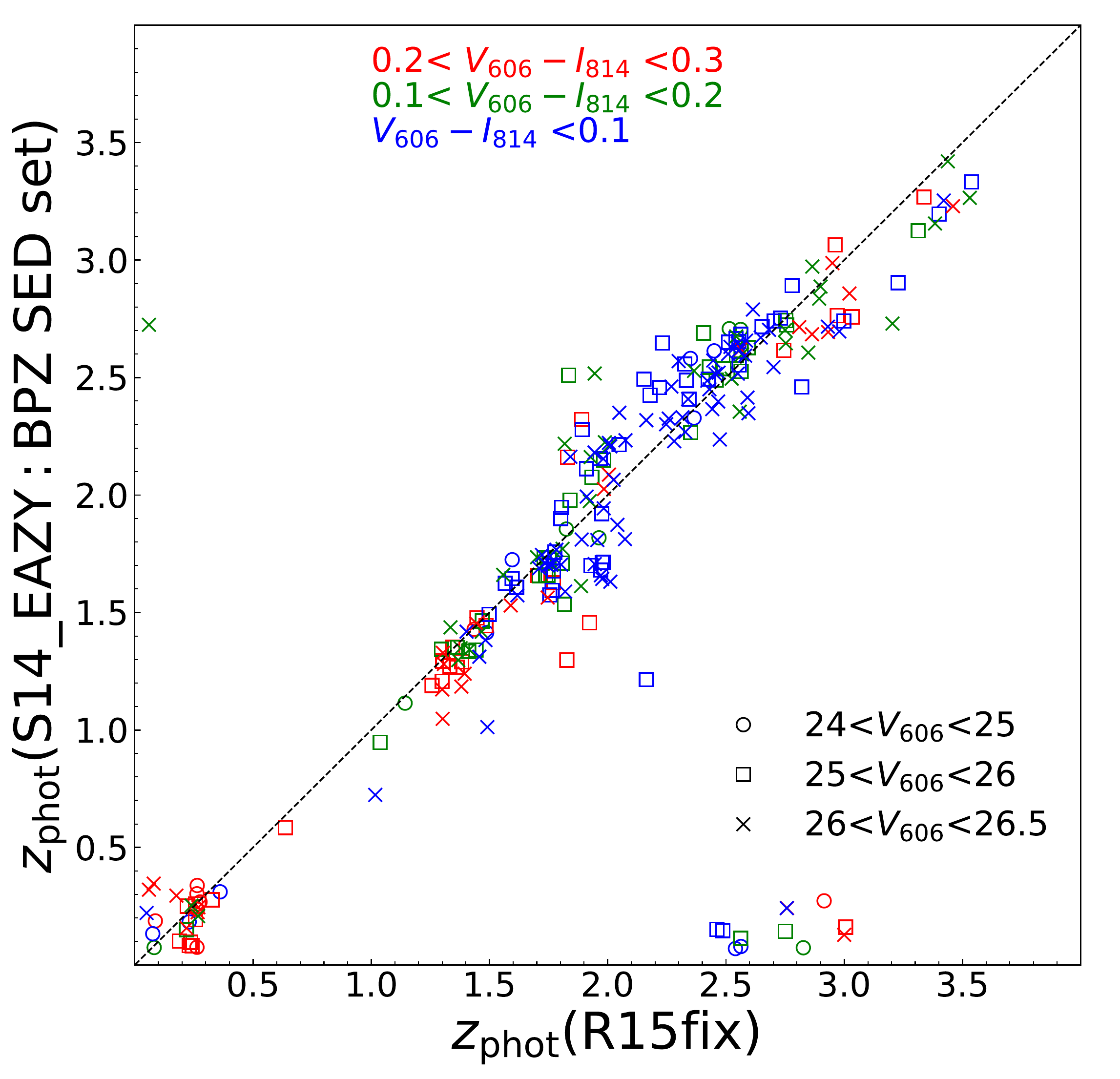}
\includegraphics[width=0.68\columnwidth]{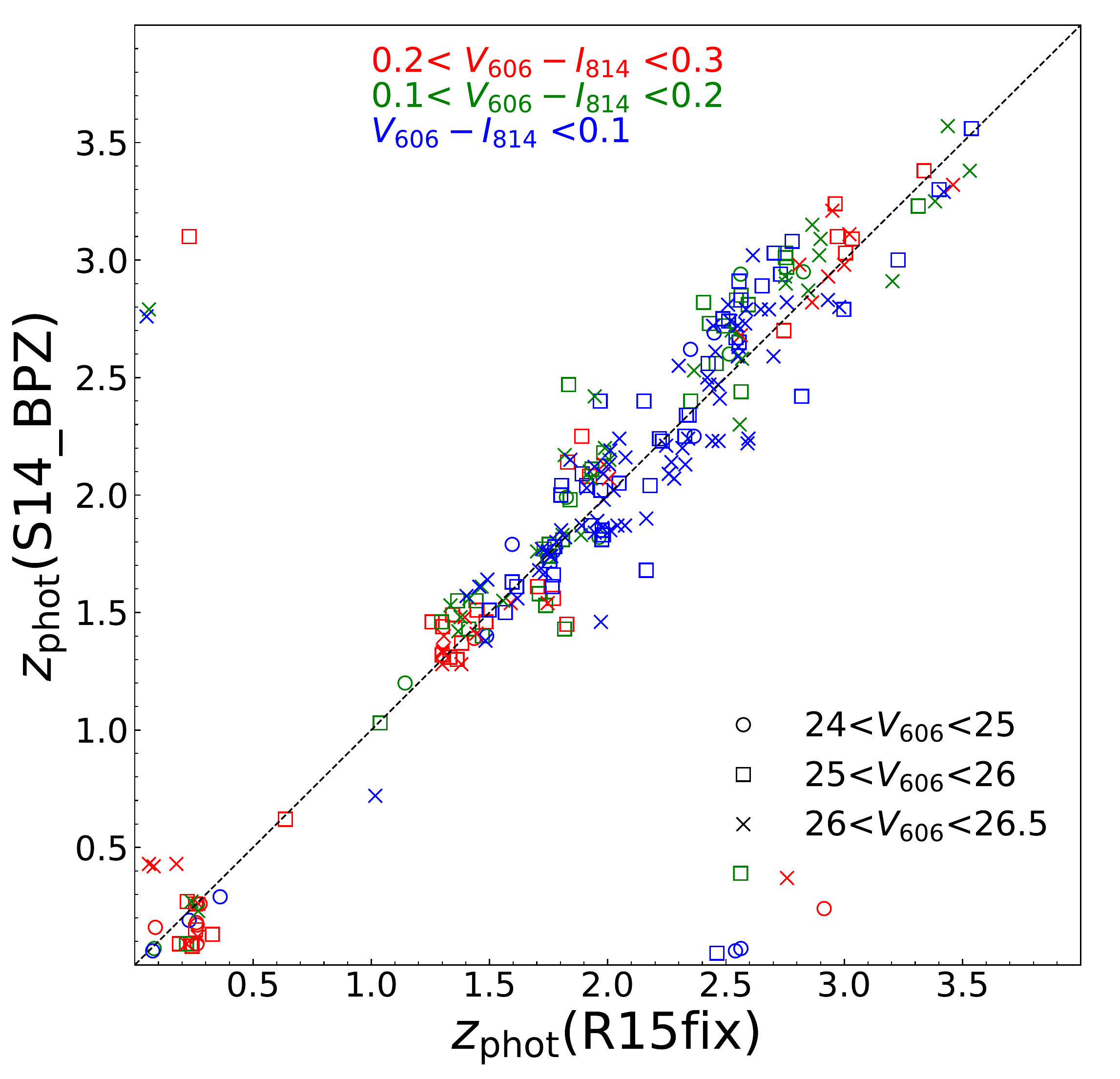}
\includegraphics[width=0.68\columnwidth]{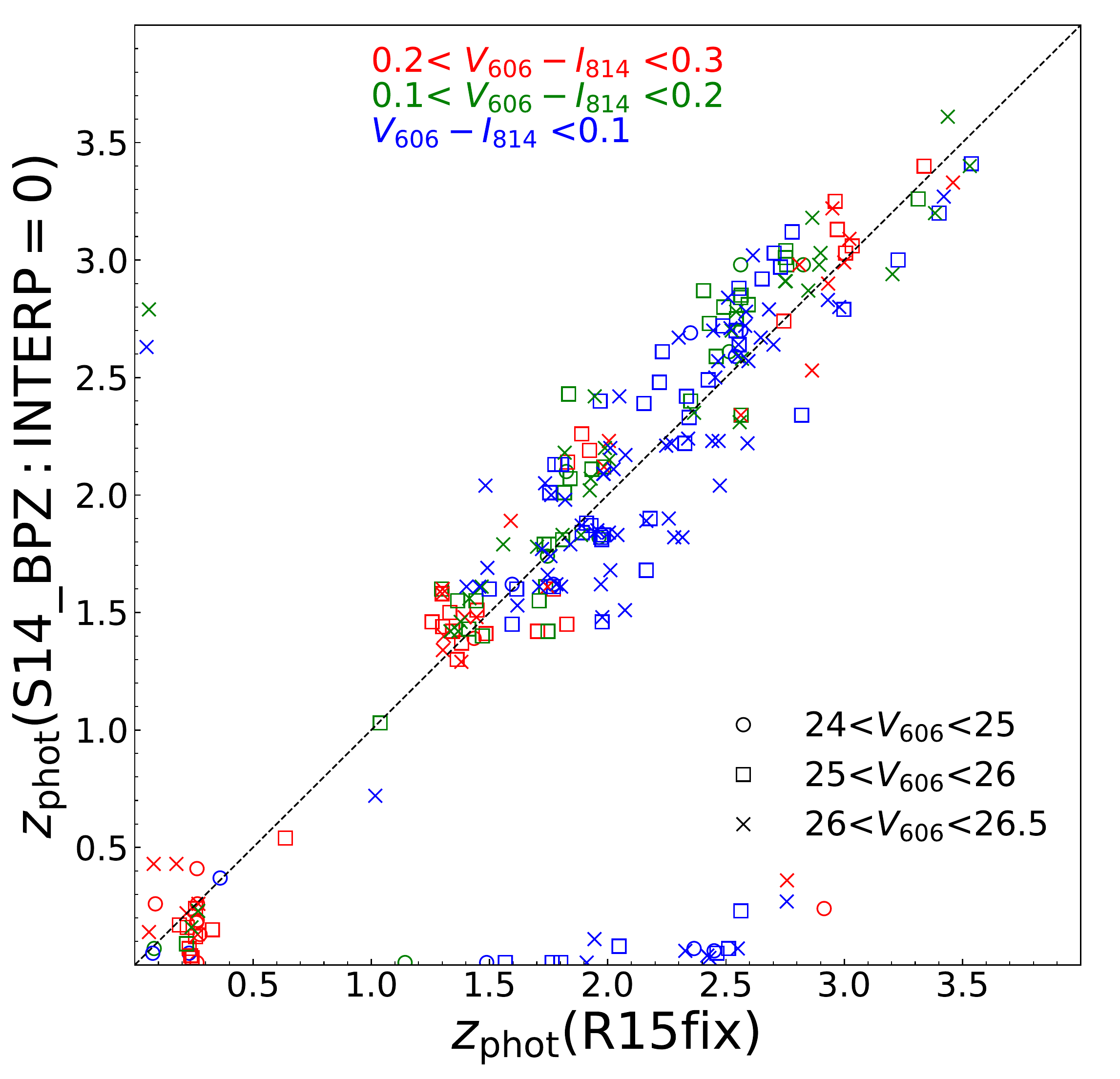}
\includegraphics[width=0.68\columnwidth]{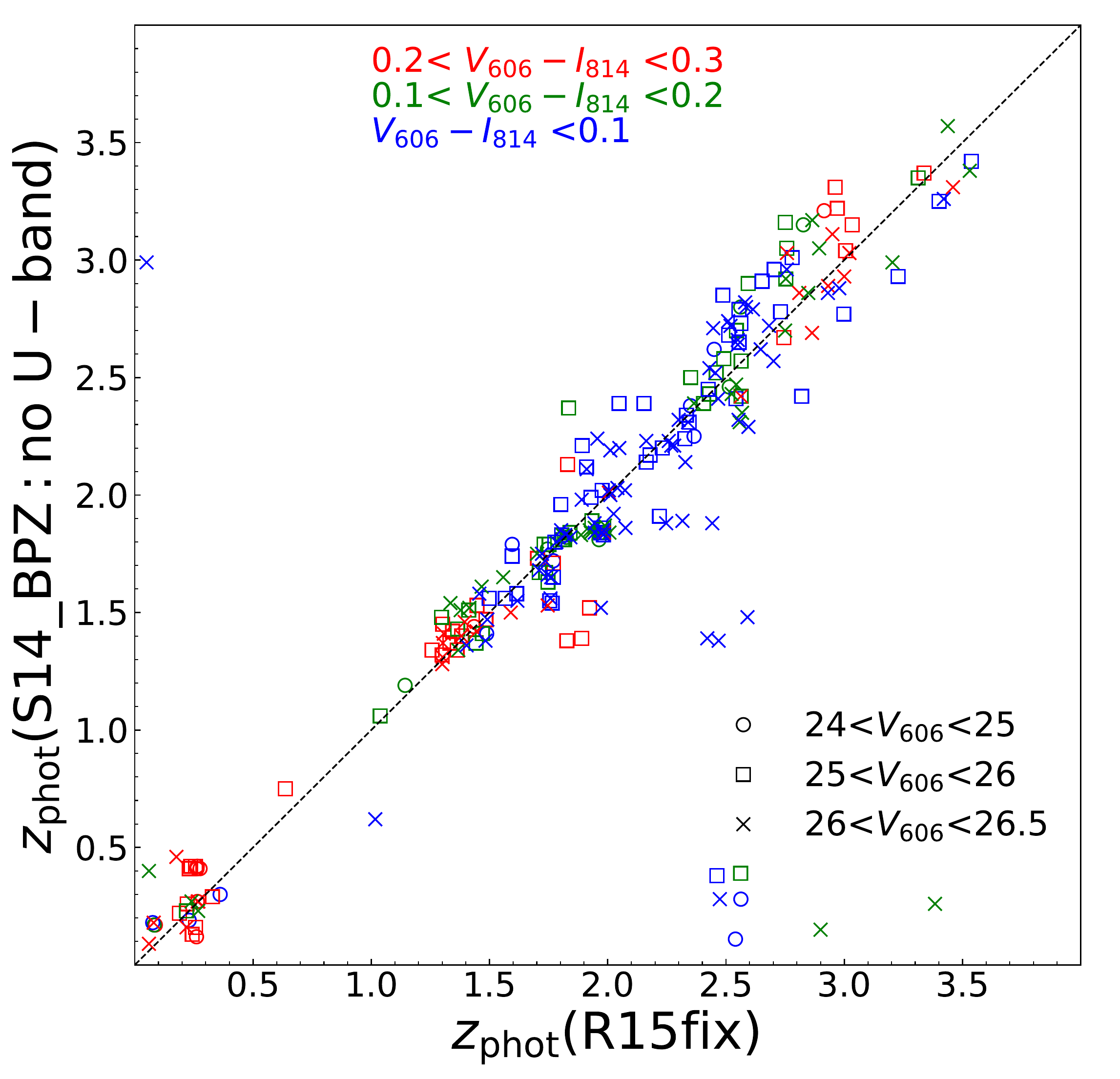}
\includegraphics[width=0.68\columnwidth]{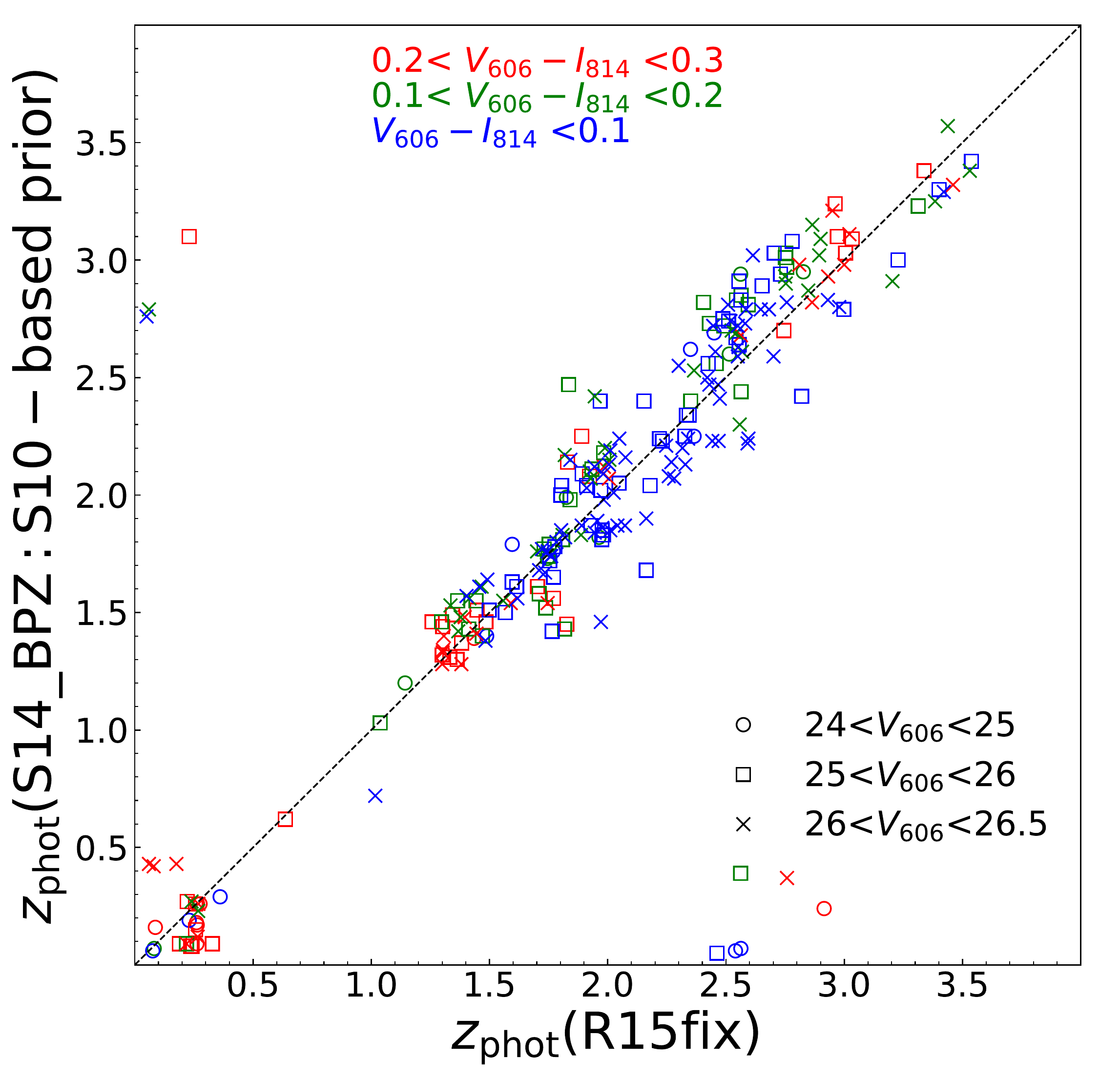}
\includegraphics[width=0.68\columnwidth]{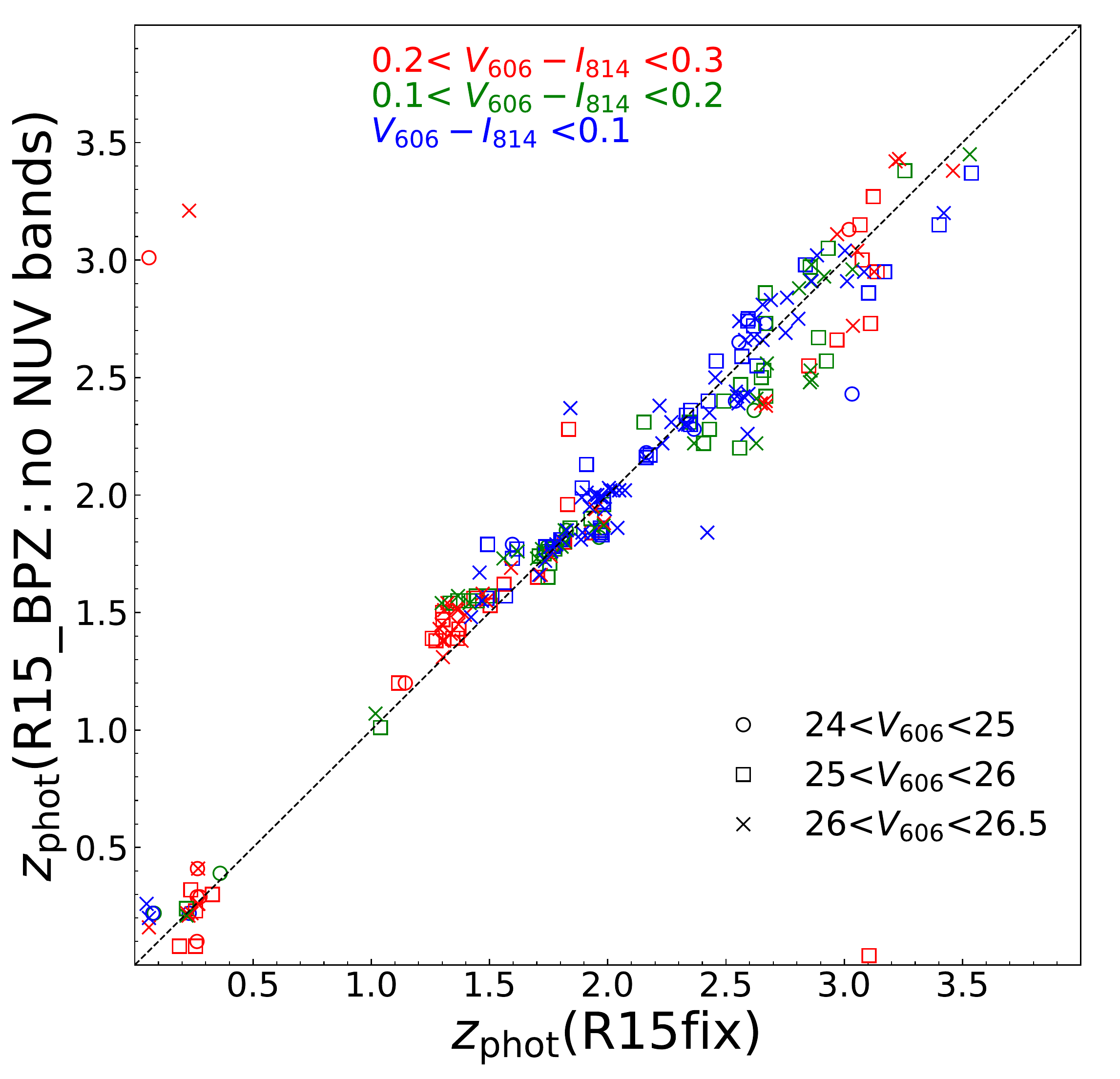}
\includegraphics[width=0.68\columnwidth]{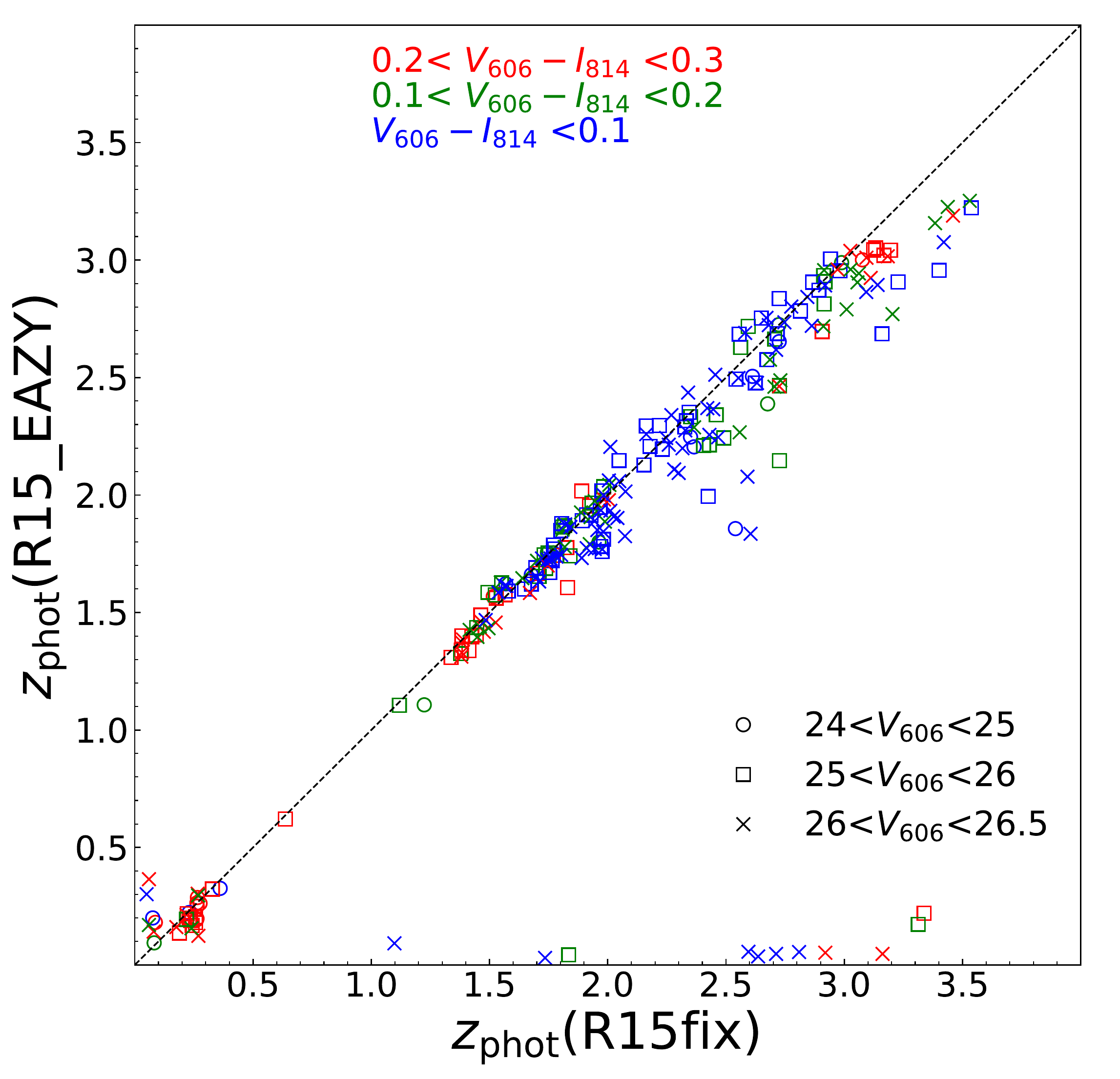}
\caption{Comparison of different sets of photo-$z$s to R15fix photo-$z$s for the colour-magnitude-selected sample, with symbols and colours similar to Figure \ref{fig:1}.
\label{fig:ccutR15fix1}}
\end{figure*}

Comparing \skeltonphotz ~to \fixphotz ~in Figure \ref{fig:ccutR15fix1} and assuming that \fixphotz ~represents the truth, shows that the \skeltonphotz ~catastrophic outliers are very asymmetric, where many galaxies with a high \fixphotz ~are assigned a low \skeltonphotz\footnote{As visible from the colour coding in Figure \ref{fig:ccutR15fix1}, these outliers occur in our full relevant range of \mbox{$V_{606}-I_{814}$} colour. Other colours are not available for the weak lensing galaxy cluster fields to which we apply our redshift calibration scheme. Therefore, we cannot follow the approach suggested by \cite{2017Speagle} to remove particularly problematic regimes in colour space.}, but not vice versa. At high redshift especially, it is obvious that the redshift distribution inferred from \skeltonphotz ~is strongly biased low. 

Now that we have established \fixphotz ~as our new calibration sample, we will calculate the relative bias of various photo-$z$s calculations and normalise it to $\langle\beta_{\mathrm{R15fix}}\rangle$. In the following subsections, we will compare the relative bias/error of the tests we have done to investigate the source of the systematic features that are present in \skeltonphotz.

\begin{table*}
\caption{Explanation of the labels.}
\label{tab:explain}
\tabcolsep=0.13cm
\begin{center}
\renewcommand{\tabularxcolumn}{m}
\begin{tabularx}{\textwidth}{l|l}
Label                     & Explanation     \\ \hline
S14\_\eazy        & The original photo-$z$s from \citetalias{Skelton20143D-HSTMASSES}. Bands used for the GOODS-South field are the ground-based $U$, $B$, $V$, \\ & $R$, $I$, $J$, $H$, $K_s$, 14 medium bands, $HST$ $F435W$, $F606W$, $F775W$, $F814W$, $F850LP$, $F125W$, $F140W$,  \\ & $F160W$ bands, and $Spitzer$ $3.6, 4.5, 5.8, 8 \mu m$ bands \tabularnewline
S14\_\eazy: $HST$ + $U$ bands & Calculated using only $HST$ bands and ground-based $U$-band \tabularnewline
S14\_\eazy: \bpz ~SED set & As S14\_\eazy: $HST$ + $U$ bands but changed the SED set from the ones employed by \citetalias{Skelton20143D-HSTMASSES} to the ones \\
& from \bpz \tabularnewline
S14\_\bpz   & Using \bpz ~with all the changes as above, with "INTERP" set to 2 (see text) \tabularnewline
S14\_\bpz: INTERP=0     & As S14\_\bpz ~but "INTERP" is set to 0 \tabularnewline
S14\_\bpz: no $U$-band    & Same as \citetalias{Skelton20143D-HSTMASSES}\_\bpz, but with $U$-band removed \tabularnewline
S14\_\bpz: S10-based prior   & Same as \citetalias{Skelton20143D-HSTMASSES}\_\bpz, but changed the prior from the default to an S10-based prior \tabularnewline
R15\_\bpz               & The original \bpz ~photo-$z$s from \citetalias{Rafelski2015UVUDF:FIELD} \tabularnewline
R15\_\bpz: no NUV bands  & Calculated using \bpz ~with the default settings (see text) using \citetalias{Rafelski2015UVUDF:FIELD} data except for the $F225W$   \\ & and $F275W$ bands removed \tabularnewline
R15\_\eazy             & The original \eazy ~photo-$z$s from \citetalias{Rafelski2015UVUDF:FIELD} \tabularnewline
\end{tabularx}
\end{center}
\end{table*}

 \begin{figure*}
 \includegraphics[width=\linewidth]{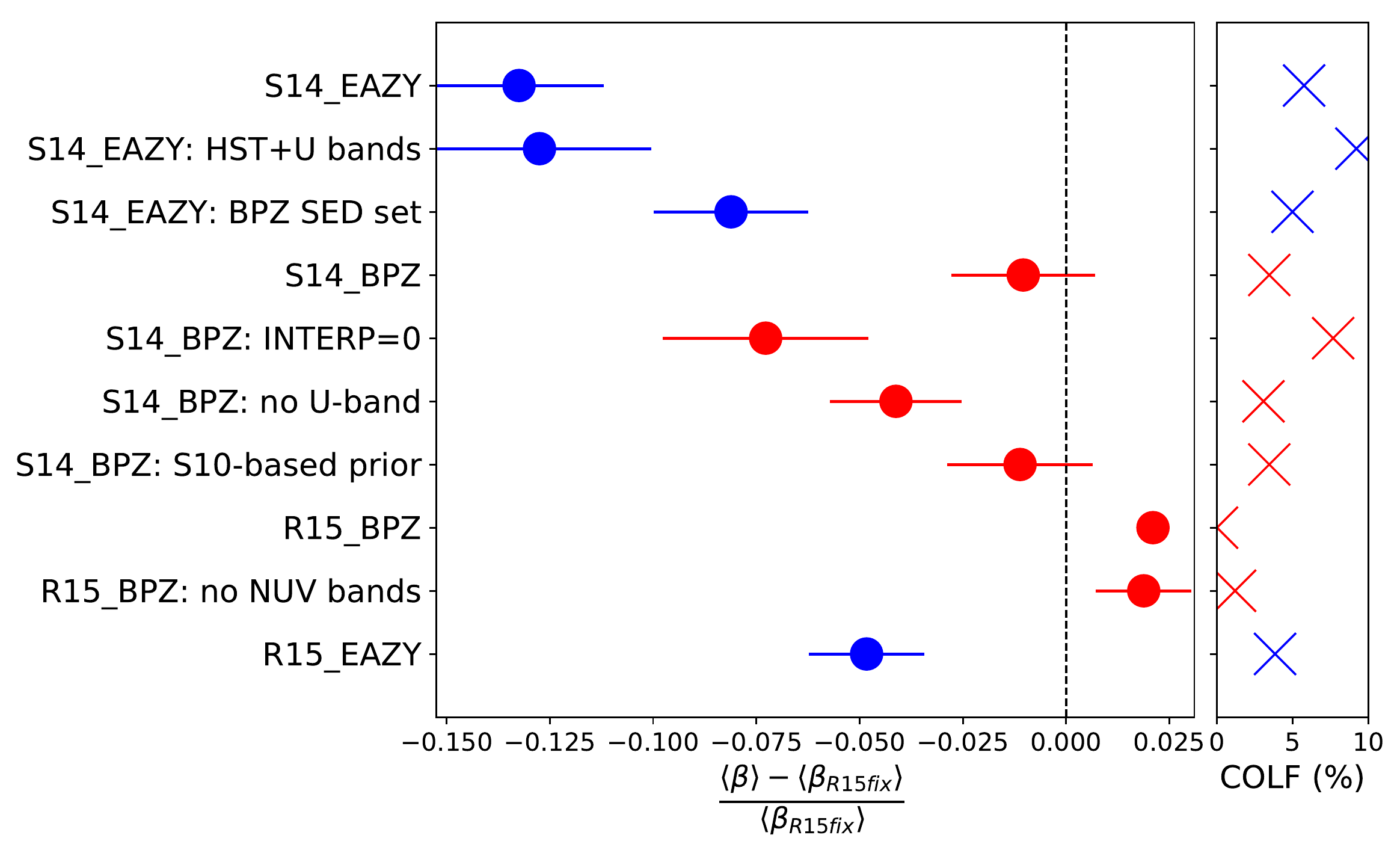}
    \caption{Relative bias in the mean geometric lensing efficiency for the colour-magnitude-selected sample and different sets of $z_{\mathrm{phot}}$ compared to \fixphotz. Description of the labels is given in Table \ref{tab:explain}. Errors are computed by bootstrapping the galaxy sample (see text), leading to hardly visible error-bars for \texttt{(R15\_BPZ)}, which differs from \texttt{(R15fix)} only because of the small redshift offsets described in Section \ref{subsection:S14R15compare}. Blue(red) indicates that the photo-$z$s are calculated using \eazy(\bpz).}
  \label{fig:vccut}
\end{figure*}

 \begin{figure*}
  \includegraphics[width=\linewidth]{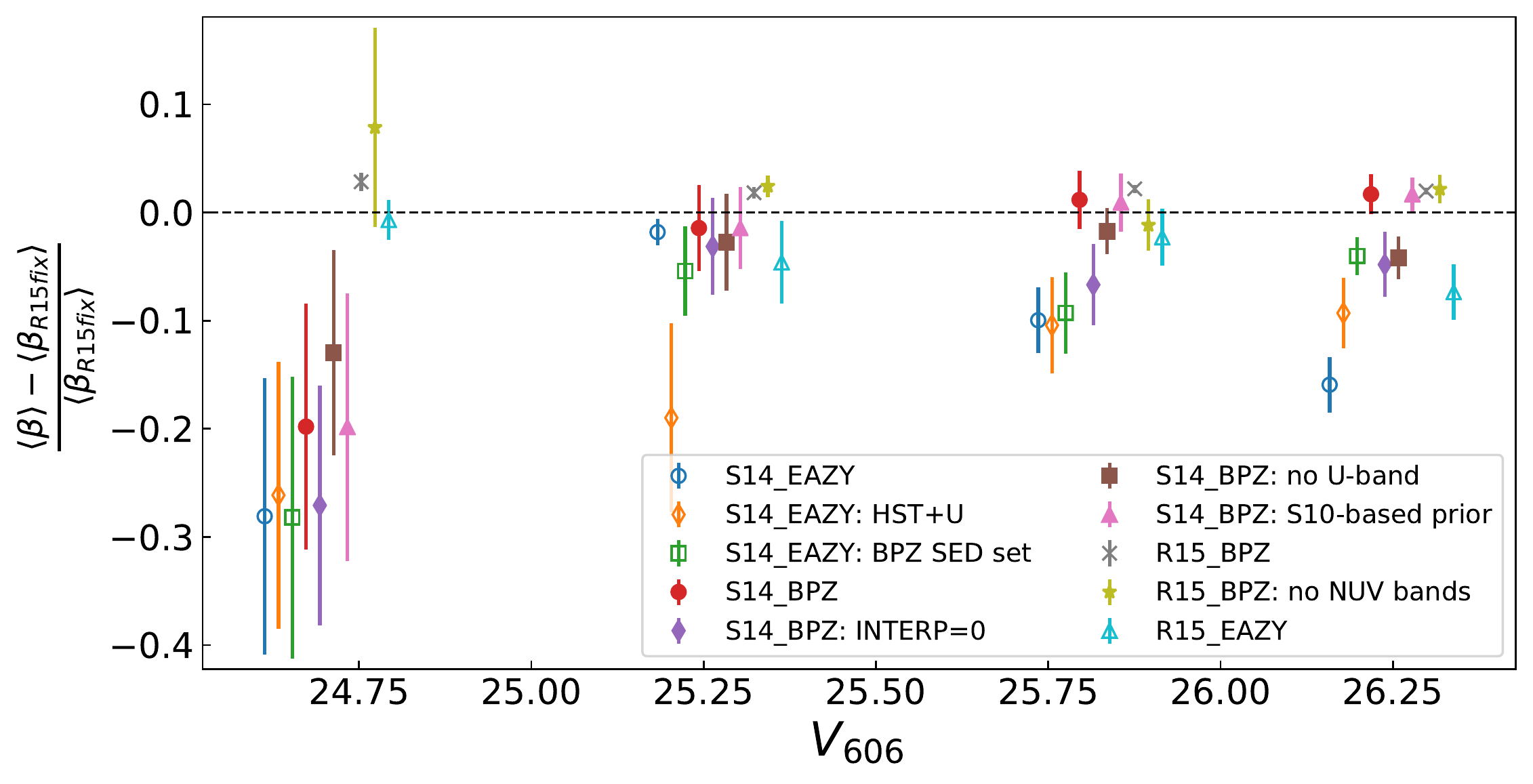}
    \caption{Same as in Figure \ref{fig:vccut} but divided into magnitude bins. The points in each magnitude bin are artificially spread out from the median value of each bin for clarity.}
  \label{fig:vccutmagbin}
\end{figure*}

\subsection{Tests using \textsc{EAZY}}

The first few tests involve photo-$z$ re-calculation using \citetalias{Skelton20143D-HSTMASSES} photometric data while employing \eazy. Figure \ref{fig:vccut} shows the relative bias/error to $\langle\beta_{\mathrm{R15fix}}\rangle$ and the corresponding COLF measured against R15fix for the colour-selected sample. We also divide our sample into $V_{\mathrm{606}}$ magnitude bins, shown in Figure \ref{fig:vccutmagbin}, to identify if there is a trend with brightness. Lastly, we show the corresponding redshift comparison plots in Figure \ref{fig:ccutR15fix1} to see if some features stand out as a function of redshift.

We find that by using only $HST$ bands and the ground-based $U$-band data from \citetalias{Skelton20143D-HSTMASSES}, we were able to reproduce the same relative bias as \citetalias{Skelton20143D-HSTMASSES} within $0.5\%$.  However, there is a higher COLF compared to \citetalias{Skelton20143D-HSTMASSES}. Referring to the top middle panel of Figure \ref{fig:ccutR15fix1}, we see that there is a slight up-scatter at \mbox{$1.8<$ \fixphotz $<2.8$} compared to the one-to-one line of mostly blue, \mbox{$V_{606}-I_{814}<0.1$} galaxies, which partially compensates the low biasing due to the increase in asymmetric COLF. The findings of this test indicate that at the wavelength range and depth of the $HST$ weak lensing data, the auxiliary data from ground-based telescope surveys, that are not as as deep as the $HST$ data especially at the longer wavelength range, do not improve the photo-$z$ determination substantially. We did an intermediate test where we removed only the IRAC data but kept the remaining ground-based data, finding that this neither improves nor worsens the relative bias. This is supported by a few studies that have stated that adding mid-IR IRAC data makes the photo-$z$s worse or does not improve them significantly \citep{Hildebrandt2010PHAT:Testing, Rafelski2015UVUDF:FIELD}. Another intermediate test that we have done is changing the prior in \eazy ~from the default $K_{\mathrm{s}}$-band to an $I$-band magnitude-based prior similar to the one employed in \bpz, finding that this changes the bias by $0.5\%$ only. We keep the $I$-band prior used here for next tests. Similarly to removing the IRAC data, we find that including or removing the ground-based $H$-band and $K_{\mathrm{s}}$-band data does not change the results significantly, again suggesting that these data may be too shallow to improve the photo-$z$s at the depth of the weak lensing data.

We then change the SED templates set from the one employed by \citetalias{Skelton20143D-HSTMASSES} to the one used in \bpz. We found that this leads to a $5\%$ improvement of the relative bias compared to \citetalias{Skelton20143D-HSTMASSES}. This also leads to smaller error bars compared to \texttt{(S14\_EAZY:HST+U)} and a slight decrease in COLF compared to \citetalias{Skelton20143D-HSTMASSES}. The redshift comparison plot (top right panel in Figure \ref{fig:ccutR15fix1}) shows that at \mbox{$25<V_{606}<26$} galaxies seem to be focused at \fixphotz ~$\sim1.9$. The relative bias decreases noticeably for the faintest magnitude bin when we change to this SED set. However, the overall relative bias of $-7.5\%$ is still not sufficient for our weak lensing analysis. 

The geometric lensing efficiency is biased at a level of $-5\%$ for $z_{\mathrm{phot}}\texttt{(R15\_EAZY)}$, which is a weaker bias than obtained for any of the \eazy ~analyses discussed above. This is also indicated by the tight correlation in the redshift comparison plot. There are fewer catastrophic outliers compared to all the tests discussed above. When investigated as function of magnitude (Figure \ref{fig:vccutmagbin}), it is worth noting that the photo-$z$s are the least biased ones in the brightest bin. 

\subsection{Tests using \textsc{BPZ}}

Up to this point, we have been unable to compute photo-$z$s that are consistent with being unbiased using \eazy. We now employ \bpz ~to re-calculate photo-$z$s. 
We first test our \bpz ~implementation using \citetalias{Rafelski2015UVUDF:FIELD} data\footnote{Here we approximate the $I_{\mathrm{814}}$ band by interpolating between $F775W$ band and $F850LP$ band from the \citetalias{Rafelski2015UVUDF:FIELD} catalogue since the band is needed for the \bpz ~prior.}. Since the SED templates that \citetalias{Rafelski2015UVUDF:FIELD} used are private, the reliability of the photo-$z$s produced in this work relies heavily on how well we can reproduce the \bpz ~photo-$z$s published by \citetalias{Rafelski2015UVUDF:FIELD}. Also, as we are using \citetalias{Skelton20143D-HSTMASSES} data that do not include NUV bands, we need to see how significantly the NUV bands affect the resulting photo-$z$ distribution.

The relative bias of \rafelskiphotz ~and the photo-$z$ set $z_{\mathrm{phot}}\texttt{(R15\_BPZ:no NUV bands)}$, which we calculate using our version of \bpz ~and \citetalias{Rafelski2015UVUDF:FIELD} data except the NUV bands, $F225W$ and $F275W$\footnote{We take $F336W$ as the equivalent to the ground-based $U$-band, so it is included in the photo-$z$ calculation.}, seems to be at a similar level ($\sim+2\%$). There is some scatter around the one-to-one line although the overall correlation is very tight. There is increased scatter at higher redshifts than at lower redshifts in the redshift comparison plot. There are two catastrophic outliers at low \fixphotz ~and one at high \fixphotz, which implies that our redshift estimates here do not produce a significant number of biased catastrophic outliers. The overall trend of \texttt{(R15\_BPZ:no NUV bands)} is very similar to \texttt{(R15\_BPZ)} in magnitude bins. We also run \bpz ~using all data and found no significant difference for the photo-$z$s with or without NUV bands. This indicates that our \bpz ~implementation is reliable and also that the NUV bands have no significant impact on the photo-$z$ determinations for our colour-magnitude-selected sample. 

Next, we calculate photo-$z$s using only $HST$ bands and the ground-based $U$-band from \citetalias{Skelton20143D-HSTMASSES}, while employing the \bpz ~SED set. The result is indicated as \texttt{S14\_BPZ} in Figures \ref{fig:ccutR15fix1}, \ref{fig:vccut} and \ref{fig:vccutmagbin}. The relative bias for this set compared to \texttt{R15fix} is found to be less than $2\%$. The COLF is low ($2\%$) and the outliers that are present are distributed more symmetrically with respect to the one-to-one line than for photo-$z$s calculated using \eazy, leading to the low bias in \betas.

The two algorithms used are very similar but they still differ in some effects, which then must be responsible for the differences. An important feature of \bpz ~is the interpolation of templates. In the previous test, we set the \texttt{INTERP} function in \bpz ~to the default value, \texttt{INTERP=2}. This means that for every adjacent SED, the code will produce two interpolated SEDs. This is likely important for fitting blue galaxies to the corresponding SEDs at high redshift, as the interpolated SEDs populate areas where there are significant gaps between the galaxy SED templates.

We find that the relative bias increases to $-7.5\%$ when we run \bpz ~with \texttt{INTERP=0}. This is similar to what we got with $z_{\mathrm{phot}}\texttt{(S14\_EAZY: BPZ SED set)}$. The trend with magnitude is very similar to that set too (Figure \ref{fig:vccutmagbin}). There is more scatter in the redshift comparison plot and a higher COLF than even compared to \citetalias{Skelton20143D-HSTMASSES}. This shows that this \texttt{INTERP} function significantly impacts the overall photo-$z$ performance. Our test for an even higher \texttt{INTERP} value reveals that there are no significant changes in the photo-$z$s. 

We also run \bpz ~on the \citetalias{Skelton20143D-HSTMASSES} data with just $HST$ bands to study the impact of ground-based $U$-band data, with the result indicated by $z_{\mathrm{phot}}\texttt{(S14\_BPZ: no U-band)}$ in Figure \ref{fig:vccutmagbin}. We find that it has a non-negligible impact on the relative bias. Removing the $U$-band tends to bias \betas ~low by $\sim-4\%$. 

Lastly, we study the impact of modifying the redshift prior on the photo-$z$s distribution. We re-calibrated the redshift prior so that the redshift prior would be more accurate in describing our magnitude-selected sample. For this we employ the $I$-band magnitude-dependent fit to the redshift distribution of COSMOS30 galaxies \citep{Ilbert2009Cosmos2} that was derived by \citet[][S10 hereafter]{Schrabback2010EvidenceCOSMOS}

\begin{equation}
p(z|i)\propto \left(\frac{z}{z_0}\right)^\alpha \left( \exp{\left[-\left(\frac{z}{z_0}\right)^\beta\right]} 
+ cu^d \exp{\left[-\left(\frac{z}{z_0}\right)^\gamma\right]}  \right), \
\label{eq:prior}
\end{equation}
where $z_0$ is computed from an assumed linear relation between the $I$-band magnitude and the median redshift, and \mbox{$u=\mathrm{max}[0,(i-23)]$}, with best-fitting parameters \mbox{$(\alpha,\beta,c,d,\gamma)=(0.678,5.606,0.581,1.464)$}. For the S10-based prior, we also choose to set the $f_t$ in \cite{Benitez2000BayesianEstimation}, which corresponds to the spectral fraction at reference magnitude $20$ for E/S0-type and Sbc/Scd-type templates to $0.05$ and $0.30$, respectively, instead of $0.35$ and $0.50$.

\begin{figure}
\includegraphics[width=\columnwidth]{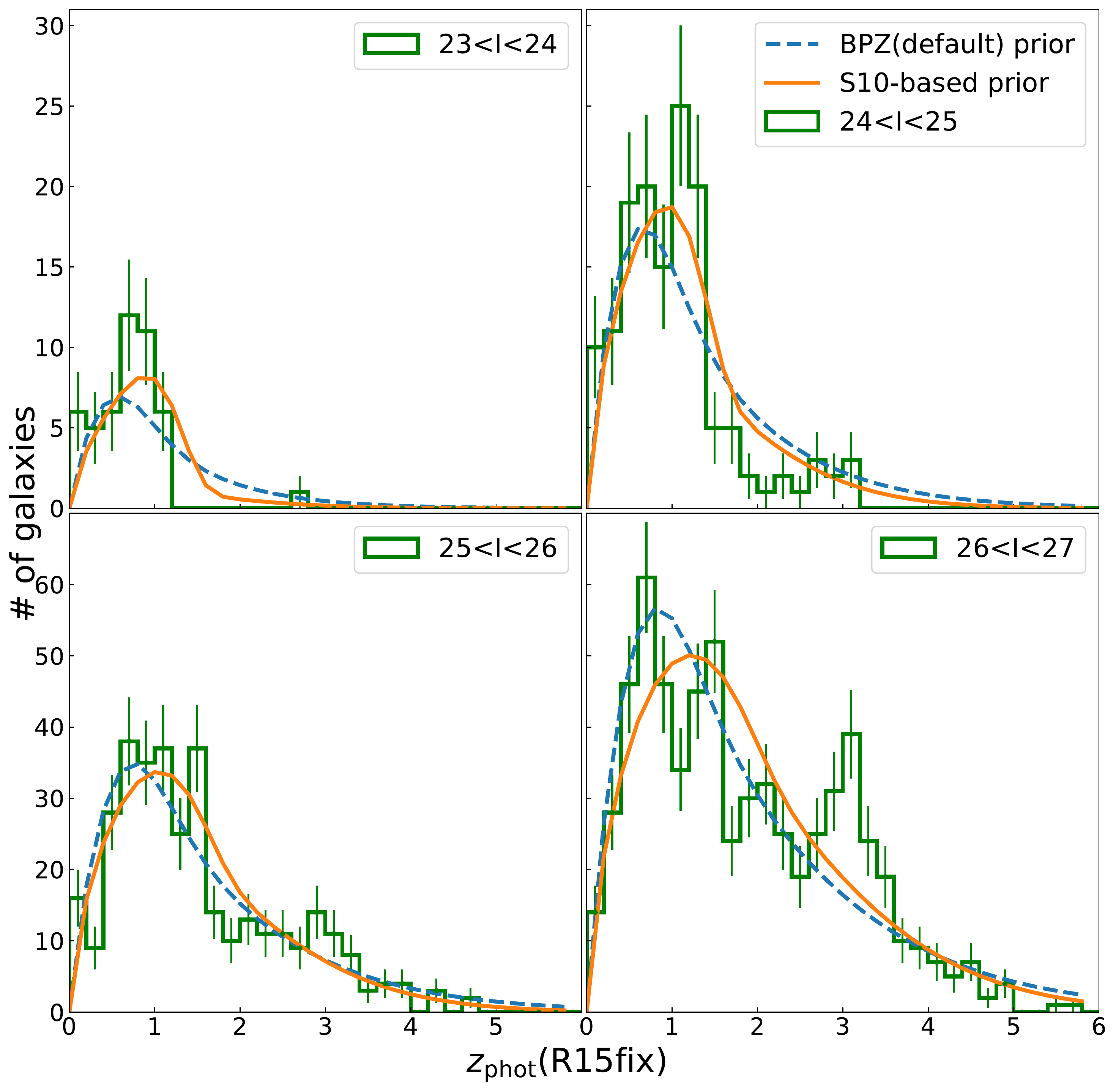}
\caption {Redshift distribution of our magnitude-selected samples using the \texttt{R15fix} photo-$z$ histogram in \textit{green}, fitted with the default prior from \bpz ~(\textit{blue dashed line}) and S10-based prior (\textit{orange solid line}). For the default prior from \bpz, we use the weighted average over the 3 types of galaxies.}
\label{fig:prior}
\end{figure}

The S10-based redshift prior fits slightly better to the \texttt{R15fix} photo-$z$s distribution of our sample than the prior from \bpz, both in terms of the median redshift difference of $0.13$ instead of $0.19$ (averaged over the magnitude bins) and the actual shape of the distribution. Figure \ref{fig:prior} illustrates the shape of the S10-based prior and compares it to \bpz ~default prior.

We find that there is not much change in the relative bias for the colour-magnitude-selected sample using the S10-based prior ($<0.5\%$) compared to $\langle\beta_{\mathrm{S14\_BPZ}}\rangle$. This indicates that the default prior in \bpz, which was calibrated using HDFN CFRS spectroscopic data, is sufficient for our redshift analysis.

\subsection{Using probability density distributions}

Past studies have suggested that using the average photometric redshift posterior probability distribution $p(z)$ of all the galaxies gives a better approximation of the true redshift distribution than using the histogram of the single-peak point-estimated photometric redshifts \citep[see e.g.][]{Heymans2012CFHTLenS:Survey,Benjamin2013CFHTLenSDistributions,Bonnett2015UsingCFHTLenS}. However, \citetalias{Schrabback2018ClusterSurvey} found that the $p(z)$ of the \citetalias{Skelton20143D-HSTMASSES} photo-$z$s cannot account for the systematic features that were identified in \ref{subsection:S14R15compare}. Similarly, we recompute \betas ~using the $p(z)$, and compare the results to the \betas ~shown in Figure \ref{fig:vccut}. We find that the bias computed using the $p(z)$ closely resembles the results from Figure \ref{fig:vccut}. For example, \texttt{(S14\_BPZ: INTERP=0)} and \texttt{(S14\_BPZ: no U-band)} still yield a low \betas ~compared to \texttt{S14\_BPZ}. Typically, \betas ~shift by $\lesssim 3\%$ only when switching from point estimates to averaged $p(z)$. Several reasons could be responsible for this behaviour, such as inaccuracies in the prior, systematic template or calibration issues, or violations of implicit Gaussian error assumptions.

\section{Accuracy of the resulting redshift calibration}
\label{section:accuracy}

In this section, we discuss the accuracy of the resulting redshift calibration. For this, we also simulate shallower fields based of \citetalias{Rafelski2015UVUDF:FIELD} photometry and estimate the resulting uncertainty from variations between CANDELS fields from the resulting \betas ~calculated using our updated photo-$z$s.

\subsection{Simulating shallower fields}
\label{subsection:simulateshallow}

\begin{figure*}
\includegraphics[width=0.7\linewidth]{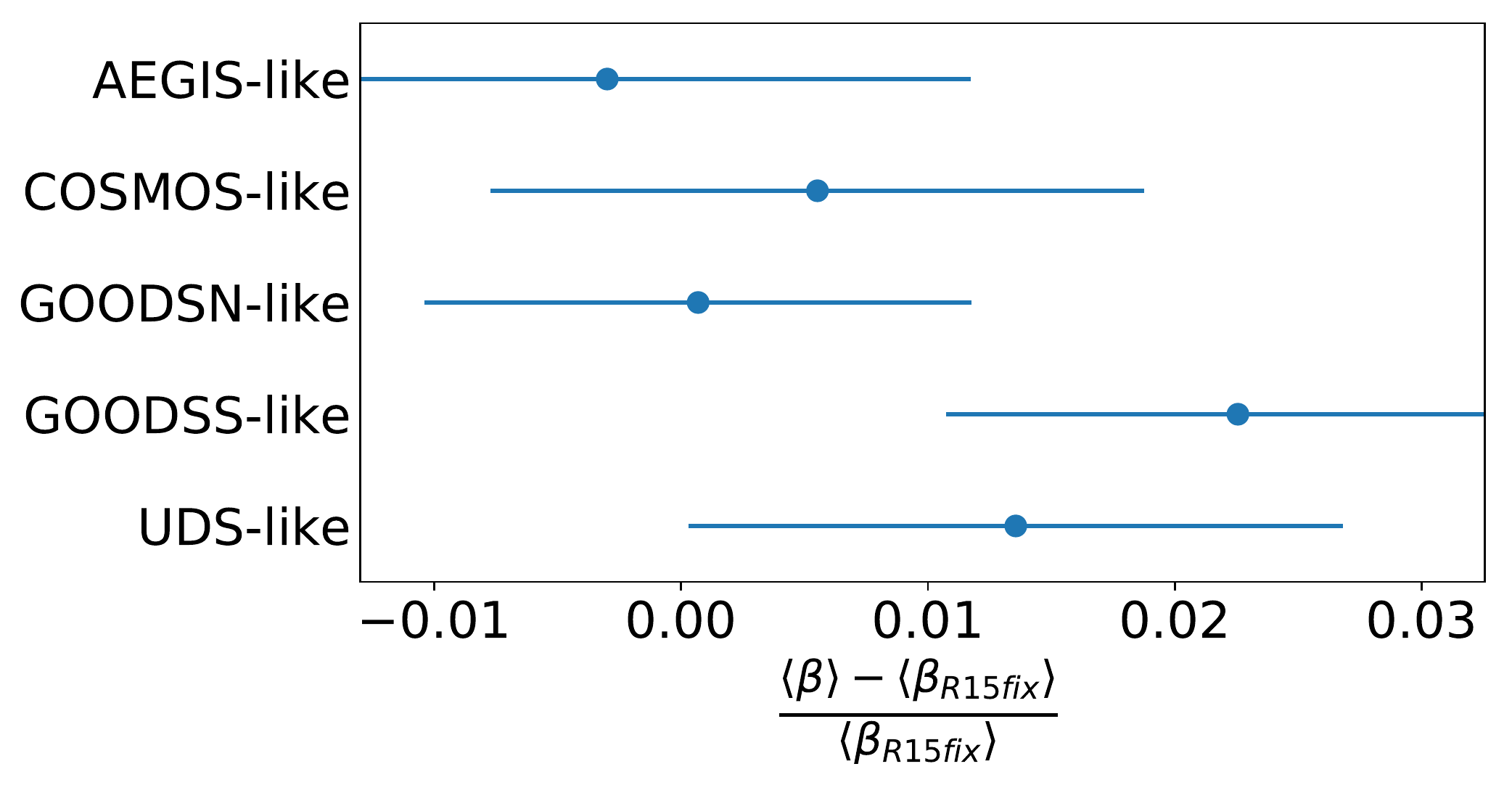}
\caption{Relative bias of the mean geometric lensing efficiency of \bpz ~photo-$z$s that we computed from \citetalias{Rafelski2015UVUDF:FIELD} photometric data after adding noise to match the depth of the different CANDELS fields.}
\label{fig:noisesim}
\end{figure*}

In this subsection, we investigate the impact of the varying noise levels in the photometric data of the different CANDELS fields on the photo-$z$ determination. One way to do this is to degrade the photometric data in deeper fields to match the noise level of the photometric data of shallower fields by adding noise. In particular, we added Gaussian noise to the \citetalias{Rafelski2015UVUDF:FIELD} photometric data such that the noise level matches the depth of the five CANDELS fields in 3D-HST. For this, the total flux in the \citetalias{Rafelski2015UVUDF:FIELD} catalogue is first converted to an aperture flux using the $F160W$ aperture-to-total ratio quoted in \citetalias{Skelton20143D-HSTMASSES}\footnote{For the \citetalias{Rafelski2015UVUDF:FIELD} galaxies that do not have a match in \citetalias{Skelton20143D-HSTMASSES}, we just use an aperture-to-total ratio of $0.7$, which is the most common value of the correction.}, then noise is added to the aperture flux. After that, these fluxes are converted back to total magnitudes using this ratio. We simulate a sufficient number of realisations of each noise level configuration. The result of the noise simulation is shown in Figure \ref{fig:noisesim}.

For the noise level of the GOODS-South field, we find a relative bias of \betas ~compared to \texttt{R15fix} of $+2.3\pm1.3\%$, which is only marginally consistent with the bias we obtained using the actual \citetalias{Skelton20143D-HSTMASSES} photometry in the overlapping HUDF area ($-1.0\pm1.7\%$, \texttt{S14\_BPZ} in Figure \ref{fig:vccut}). However, it agrees well with the $+2.1\%$ bias obtained for \texttt{R15\_BPZ} (see Figure \ref{fig:vccut}). A cause for this might be the difference in photometric zero point offset calculation between \citetalias{Skelton20143D-HSTMASSES} and \citetalias{Rafelski2015UVUDF:FIELD}. In the most extreme case, we found that in the $F435W$-band, the average magnitude offset is $-0.2$ mag in the \citetalias{Skelton20143D-HSTMASSES} photometric catalogue compared to the \citetalias{Rafelski2015UVUDF:FIELD} catalogue. The zero point offset is also inherited by the noise-added simulations, therefore systematically biasing the \betas ~the same way. 

As visible in Figure \ref{fig:vccut}, \texttt{R15\_BPZ} leads to a bias of $+2.1\%$. As the average over all five noise configuration (see Figure \ref{fig:noisesim}) we find a small positive relative bias of $+0.8\pm1.0\%$. This implies that the noise leads to a bias of $-1.3\pm1.0\%$, which is consistent with \texttt{S14\_BPZ}. This is compensated by the relative bias of \betas ~for \texttt{R15fix} compared to $z_{\mathrm{spec/grism}}$ (see Figure \ref{fig:nmad1}), which is $1.2\pm0.7\%$. Combining these two together we expect a total relative bias of $0.1\pm1.4\%$ for photo-$z$s computed with the \texttt{S14\_BPZ} setup.

\subsection{Accounting for variation between the CANDELS fields}
\label{subsection:accountingfieldvariation}
We then re-calculate the photo-$z$s using \citetalias{Skelton20143D-HSTMASSES} photometric data and our \texttt{S14\_BPZ} setup for all five CANDELS fields. We use additional $G,B,I$, and $Z$-bands data from the ground for some of the CANDELS fields to supplement the absence of $F435W$\footnote{We checked that using $G$ or $B$-band as a substitute for the $F435W$-band does not affect the photo-$z$s in a significant way.}, $F775W$, and $F850LP$ data. The summary of the bands used in the re-calculation of photo-$z$s is as follows, where we refer the reader to \citetalias{Skelton20143D-HSTMASSES} regarding details on the individual bands:

\begin{itemize}
\item AEGIS: $U$, $G$, $I$, $Z$, $F606W$, $F814W$, $F125W$, $F140W$, $F160W$
\item COSMOS: $U$, $B$, $I$, $Z$, $F606W$, $F814W$, $F125W$, $F140W$, $F160W$
\item GOODSN: $U$, $F435W$, $F606W$, $F775W$, $F850LP$, $F125W$, $F140W$, $F160W$
\item GOODSS: $U$, $F435W$, $F606W$, $F775W$, $F814W$, $F850LP$, $F125W$, $F140W$, $F160W$
\item UDS: $U$, $B$, $I$, $Z$, $F606W$, $F814W$, $F125W$, $F140W$, $F160W$.
\end{itemize}

\begin{figure*}
\includegraphics[width=0.7\linewidth]{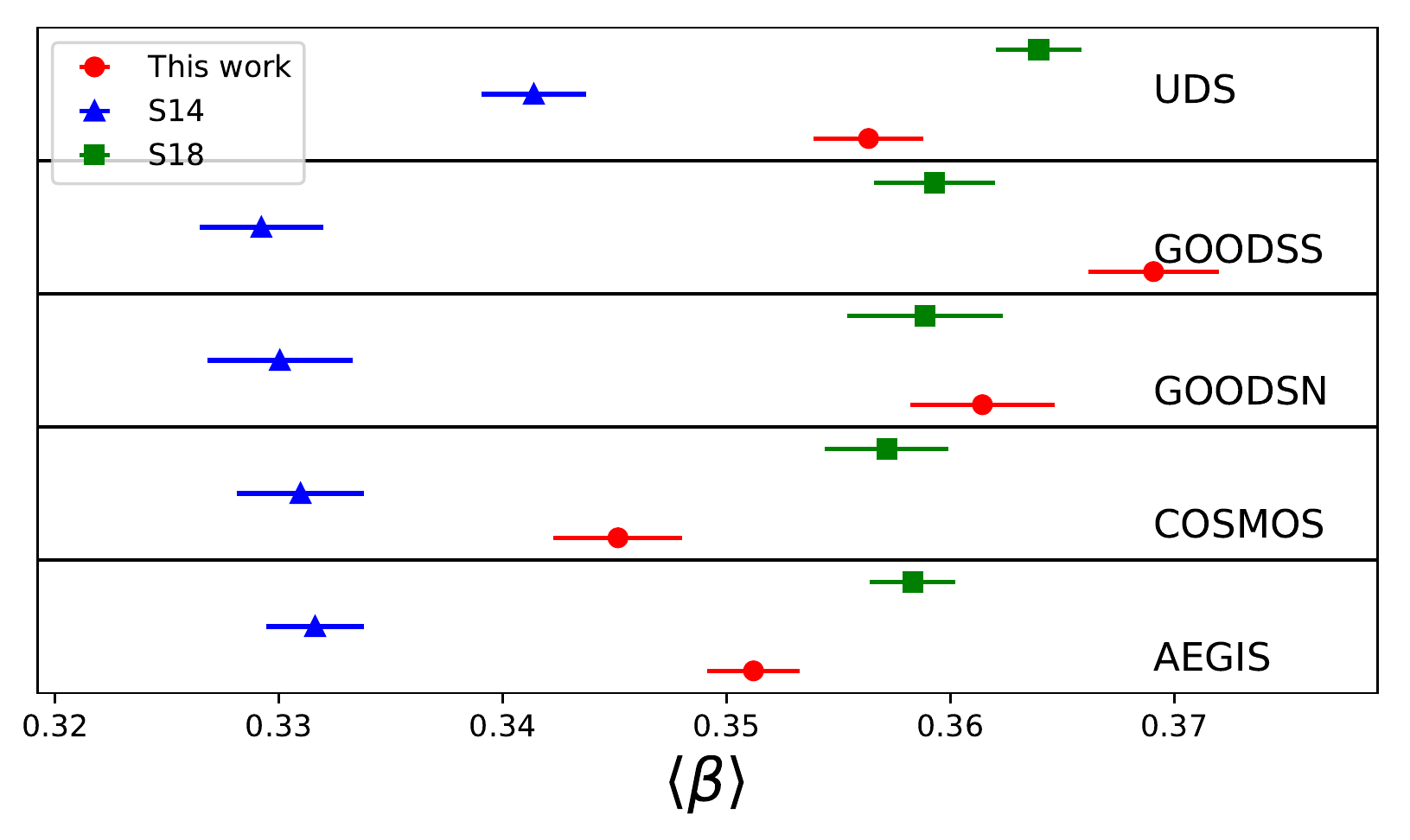}
\caption{Mean geometric lensing efficiency computed for the CANDELS fields using the original \citetalias{Skelton20143D-HSTMASSES} catalogues, the empirically corrected catalogues from \citetalias{Schrabback2018ClusterSurvey} and our new catalogues computed using the \texttt{S14\_BPZ} setup.
\label{fig:redos14}}
\end{figure*}

Here we also compare our \betas ~to the ones computed by \citetalias{Schrabback2018ClusterSurvey} using their statistical correction to the photo-$z$s (see Figure \ref{fig:redos14}). The mean \betas ~of the five CANDELS fields from our work is $0.3566\pm0.0092$ which is consistent with the estimate $0.3595\pm0.0026$ from the \citetalias{Schrabback2018ClusterSurvey}-corrected catalogues. Here the uncertainty on the mean is computed from the variation between the five fields, corresponding to a $2.6\%$ relative uncertainty for our results (the correction from \citetalias{Schrabback2018ClusterSurvey}). A part of the variation between the different fields comes from large-scale structure variations. This does, however, not explain the larger scatter for our results. 

Naively we would expect the opposite behaviour, as \citetalias{Schrabback2018ClusterSurvey} apply the same empirical redshift correction to all five CANDELS fields ignoring their variation in depth, which is in principle accounted for in our analysis. A possible explanation for the observed behaviour may be given by the fact that we include fewer bands in our analysis. This can lead to an increased scatter in the \betas ~between the five fields in two ways. First, fewer bands increase the impact of redshift focussing effects, which can differ between the different fields as they are not covered in exactly the same filters and with the same depth. Second, residual photometric calibration errors have a bigger impact on the photo-$z$s if fewer bands are used, whereas their impact averages out more if a larger number of filters is available. 

Within our analysis framework we are not able to correct for these effects and therefore include the $2.6\%$ relative uncertainty in our systematic error budget. Added in quadrature to the uncertainty estimated in Sections \ref{subsection:S14R15compare} and \ref{subsection:simulateshallow} this yields a total relative systematic uncertainty of the \betas ~calibration of $3.0\%$.

\subsection{Update to the S18 cluster masses}

\begin{figure*}
  \includegraphics[width=1.5\columnwidth]{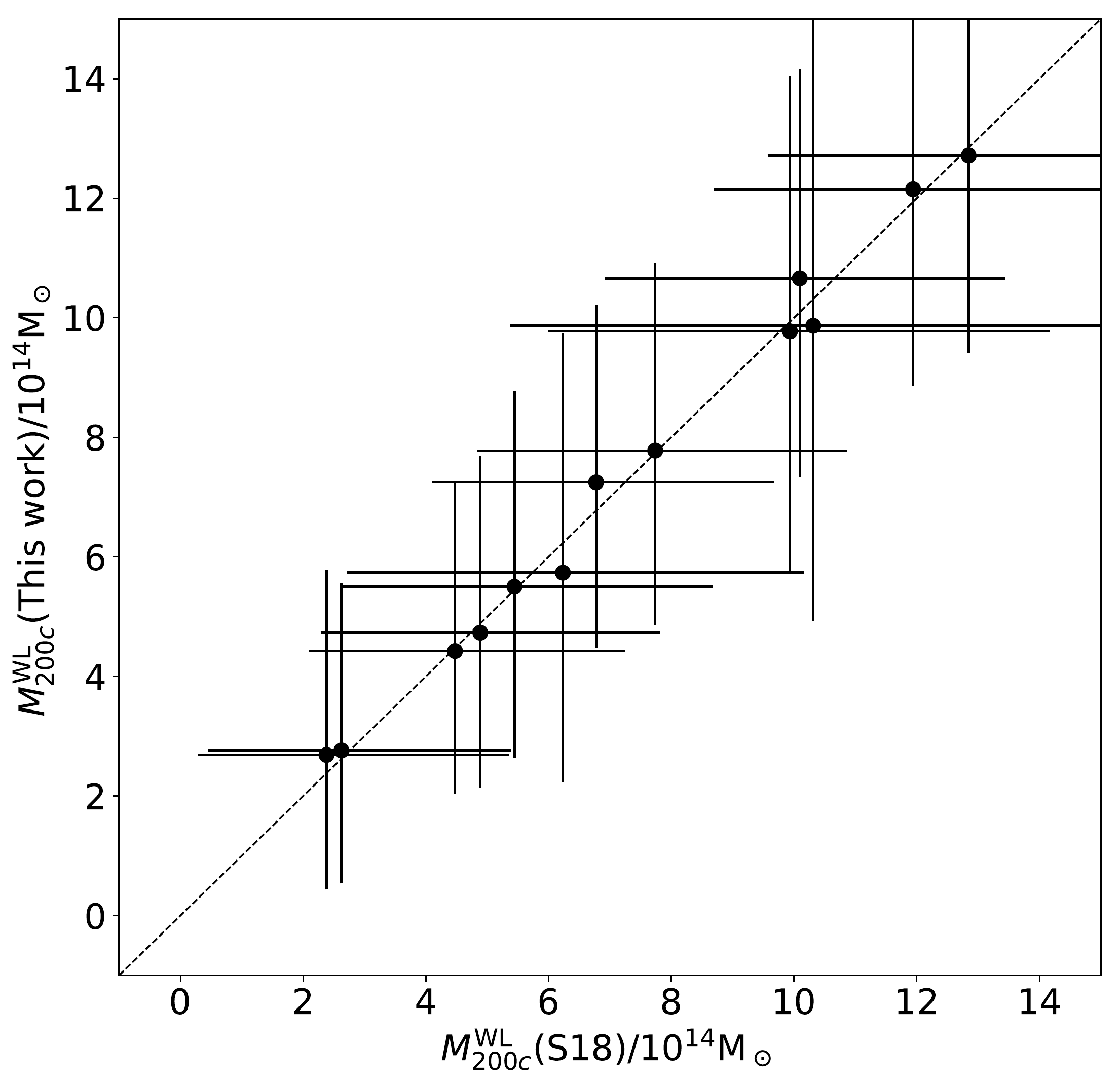}
\caption{Comparison of weak lensing mass estimates using the X-ray centres of this work versus \citetalias{Schrabback2018ClusterSurvey}. The mass constraints are based on NFW fits to the reduced shear profiles using scales \mbox{$500\mathrm{\,kpc}<r<1.5$} Mpc employing the \citet{Diemer2015ACONCENTRATIONS} $c(\mathrm{M})$ relation for over-densities in $200\mathrm{c}$, and correcting for mass modelling bias as done by \citetalias{Schrabback2018ClusterSurvey}. All errors are statistical $68\%$ uncertainties, including the contributions from shape noise, uncorrelated large-scale, and  line-of-sight variations in the redshift distribution. See \citetalias{Schrabback2018ClusterSurvey} for further details on the weak lensing mass measurements.}
\label{fig:mass}
\end{figure*}

Using our new \bpz ~CANDELS photo-$z$ catalogues for the redshift calibration we recompute the weak lensing masses of the high-redshift SPT-SZ clusters from \citetalias{Schrabback2018ClusterSurvey} and compare them to the original estimates in Figure \ref{fig:mass}. 

We find that the resulting mass estimates of the clusters in this work are very consistent with the mass estimates from \citetalias{Schrabback2018ClusterSurvey}: compared to \citetalias{Schrabback2018ClusterSurvey} the masses shift by $+1\%$ on average. 

\section{Conclusions}
\label{section:conclude}

Firstly, through comparison with $z_{\mathrm{spec/grism}}$, we have established that very deep photo-$z$s from \citetalias{Rafelski2015UVUDF:FIELD} constitute a good calibration sample that solves the problem of incomplete spec-$z$s for faint, high-redshift galaxies. Here we applied a small bias correction to the \citetalias{Rafelski2015UVUDF:FIELD} photo-$z$s, and denoted these corrected photo-$z$s as \texttt{R15fix}.

When comparing the \citetalias{Rafelski2015UVUDF:FIELD} and \citetalias{Skelton20143D-HSTMASSES} photo-$z$s, we found that \citetalias{Skelton20143D-HSTMASSES} suffers from systematic features, most importantly catastrophic outliers, which systematically bias the distribution of the photo-$z$s low. This bias of the photo-$z$s is problematic for weak lensing studies as biased photo-$z$s will lead to a biased interpretation of the weak lensing signal. For our colour-magnitude-selected sample, the relative bias in \betas ~of the \citetalias{Skelton20143D-HSTMASSES} photo-$z$s compared to \fixphotz ~is $-13.2\%$.

In general, the absolute value of a bias is less of concern since it can be compensated in a cosmological analysis. Instead, it is the accuracy with which the bias can be determined that propagates into the systematic error budget of the cosmological constraints. In order to better constrain this accuracy, we have studied the cause of the systematic features by re-calculating photo-$z$s to test the impact of differences in the analysis and the data between the \citetalias{Skelton20143D-HSTMASSES} and \citetalias{Rafelski2015UVUDF:FIELD} photo-$z$s. We have found that, although the \citetalias{Skelton20143D-HSTMASSES} data have lower S/N compared to \citetalias{Rafelski2015UVUDF:FIELD}, we are able to achieve a low relative bias of less than $2\%$ by using \bpz ~instead of \eazy. Apart from changing the SED set, we found that the interpolation of the SED set as implemented in \bpz ~has the biggest impact on the relative bias for the colour-magnitude-selected sample. We also found that the inclusion of $U$-band data from ground-based telescopes is crucial to obtain accurate photo-$z$ distributions. FIR IRAC data and the other ground-based data only have a small impact on the relative bias.

For \bpz ~we also tested the use of an alternative prior based on COSMOS-30 photo-$z$s, finding that it has only a minor impact.
Using \eazy ~we tried to match the \bpz ~properties, employing the same templates and priors. Nevertheless, we have been unable to obtain unbiased results with \eazy. Using the averaged probability density distribution of the photo-$z$s instead of using single-peak point-estimated photo-$z$s did not change the bias results.

We investigated the impact of noise by degrading the \citetalias{Rafelski2015UVUDF:FIELD} HUDF photometry to the depth of the different CANDELS fields. Combining these results with the estimates from the initial spectroscopic comparison we expect that our setup running \bpz ~on \citetalias{Skelton20143D-HSTMASSES} data should yield unbiased estimates of \betas ~with a systematic uncertainty of $1.4\%$. Using this setup we then recomputed the photo-$z$s for all five CANDELS fields. Here we detected a larger field-to-field variation in the \betas ~compared to \citetalias{Schrabback2018ClusterSurvey}, which may be caused by the inclusion of fewer bands in our analysis, leading to a total systematic uncertainty of the \betas ~calibration of $2.9\%$. Using our updated CANDELS catalogues as reference sample we recomputed the cluster mass estimates from \citetalias{Schrabback2018ClusterSurvey}, finding an average increase of the masses by $+1\%$.

In the future, we will apply these updated photo-$z$s in combination with an updated shear calibration (Hern\'{a}ndez-Mart\'{i}n et al. (submitted)) and extended $HST$ data sets, e.g. from the SPT ACS snapshot survey (Schrabback et al. in prep) to further improve the mass calibration of high-redshift galaxy clusters.

\section*{Acknowledgements}

This work is based on observations made with the NASA/ESA {\it Hubble Space Telescope}, using imaging data from the $SPT$ follow-up GO programmes 12246 (PI: C.~Stubbs) and 12477  (PI: F.~W.~High), as well as archival data from
GO programmes 9425, 9500, 9583, 10134, 12064, 12440, and 12757, obtained via the data archive at the Space
Telescope Science Institute, and catalogues based on observations taken by the 3D-HST Treasury Program (GO 12177 and 12328) and the UVUDF Project (GO 12534, also based on data from GO programmes 9978, 10086, 11563, 12498).
STScI is operated by the Association of Universities for Research in Astronomy, Inc. under NASA contract NAS 5-26555.
The MUSE data is supported by the ERC advanced grant 339659-MUSICOS (R. Bacon) and based on observations made with ESO telescopes at the La Silla Paranal Observatory under program ID 60.A-9100(C). SFR, TS, and DA acknowledge support from the German Federal Ministry of Economics and Technology (BMWi) provided through DLR under projects 50 OR 1210, 50 OR 1308, 50 OR 1407, 50 OR 1610 and 50 OR 1803. SFR also acknowledges the financial support from the DAAD Abschlussstipendium funding. SFR is a member of and received financial support for this research from the International Max Planck Research School (IMPRS) for Astronomy and Astrophysics at the Universities of Bonn and Cologne. HH is supported by a Heisenberg grant of the Deutsche Forschungsgemeinschaft (Hi 1495/5-1) as well as an ERC Consolidator Grant (No. 770935). We thank Beatriz Hern\'{a}ndez-Mart\'{i}n, Hannah Zohren and Nils Weissgerber for useful discussions. We also thank Henk Hoekstra and Maurilio Pannella for useful comments on the manuscript. We would also like to thank the anonymous reviewer for the con-structive comments.

\section*{Data availability}
The data underlying this study are available in the links within the article. The photo-$z$ catalogues generated are available on request to the corresponding author, SFR.

\bibliographystyle{mnras}
\bibliography{mnraspaper.bib}

\appendix
\section{Result for a purely magnitude-selected sample}
\label{app:magselect}

\begin{figure*}
 \includegraphics[width=0.68\columnwidth]{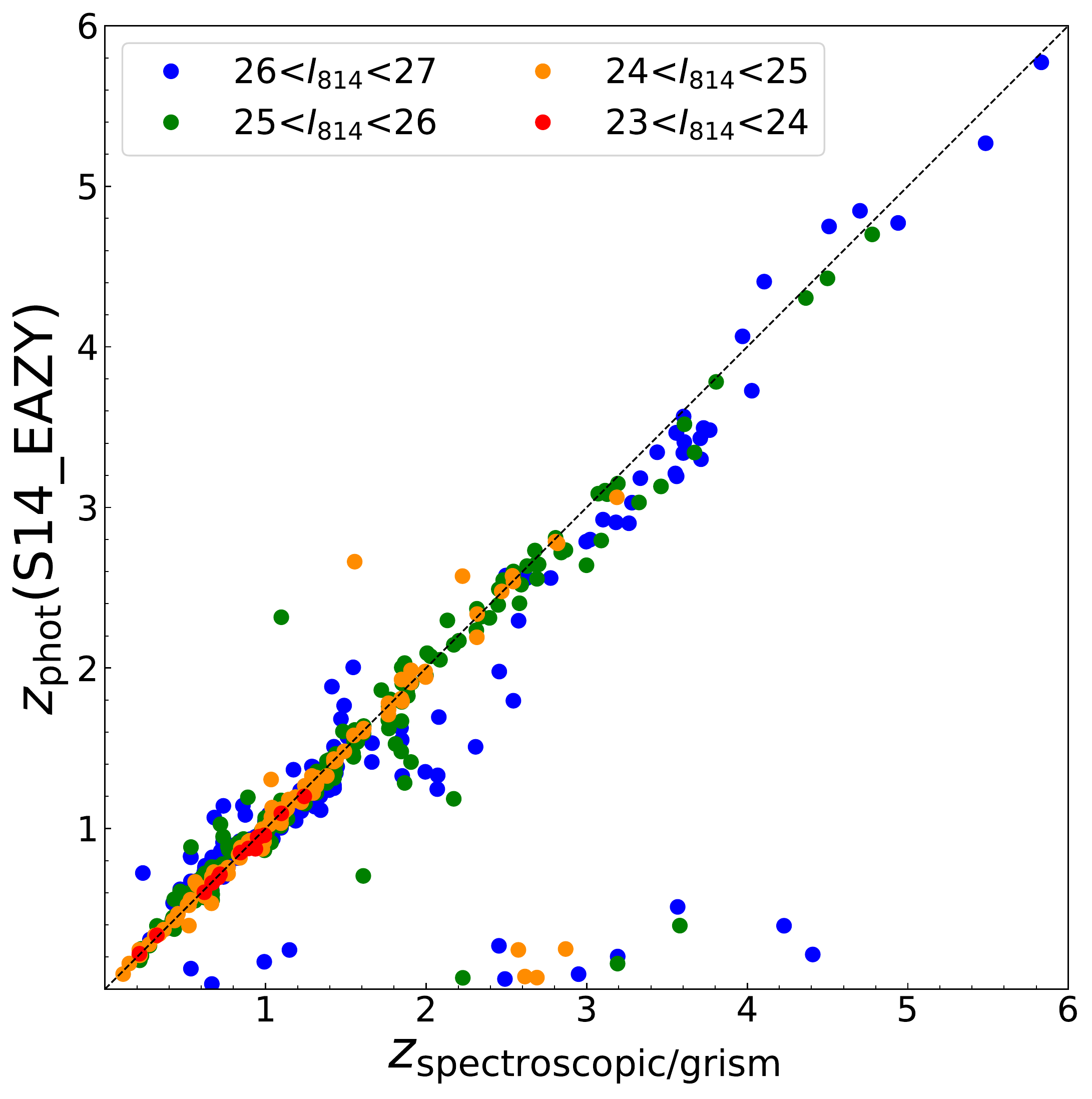}
 \includegraphics[width=0.68\columnwidth]{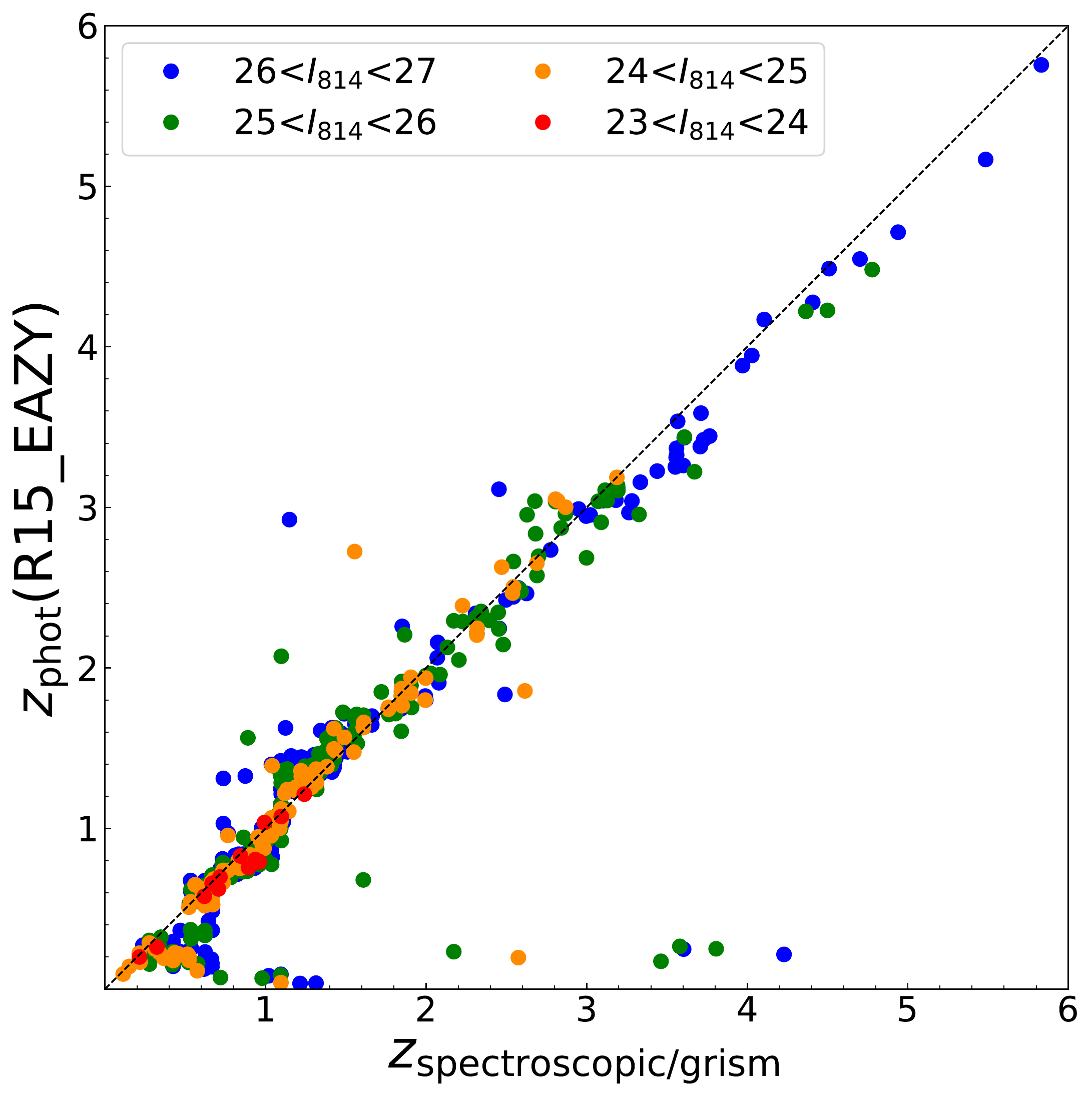}
  \includegraphics[width=0.68\columnwidth]{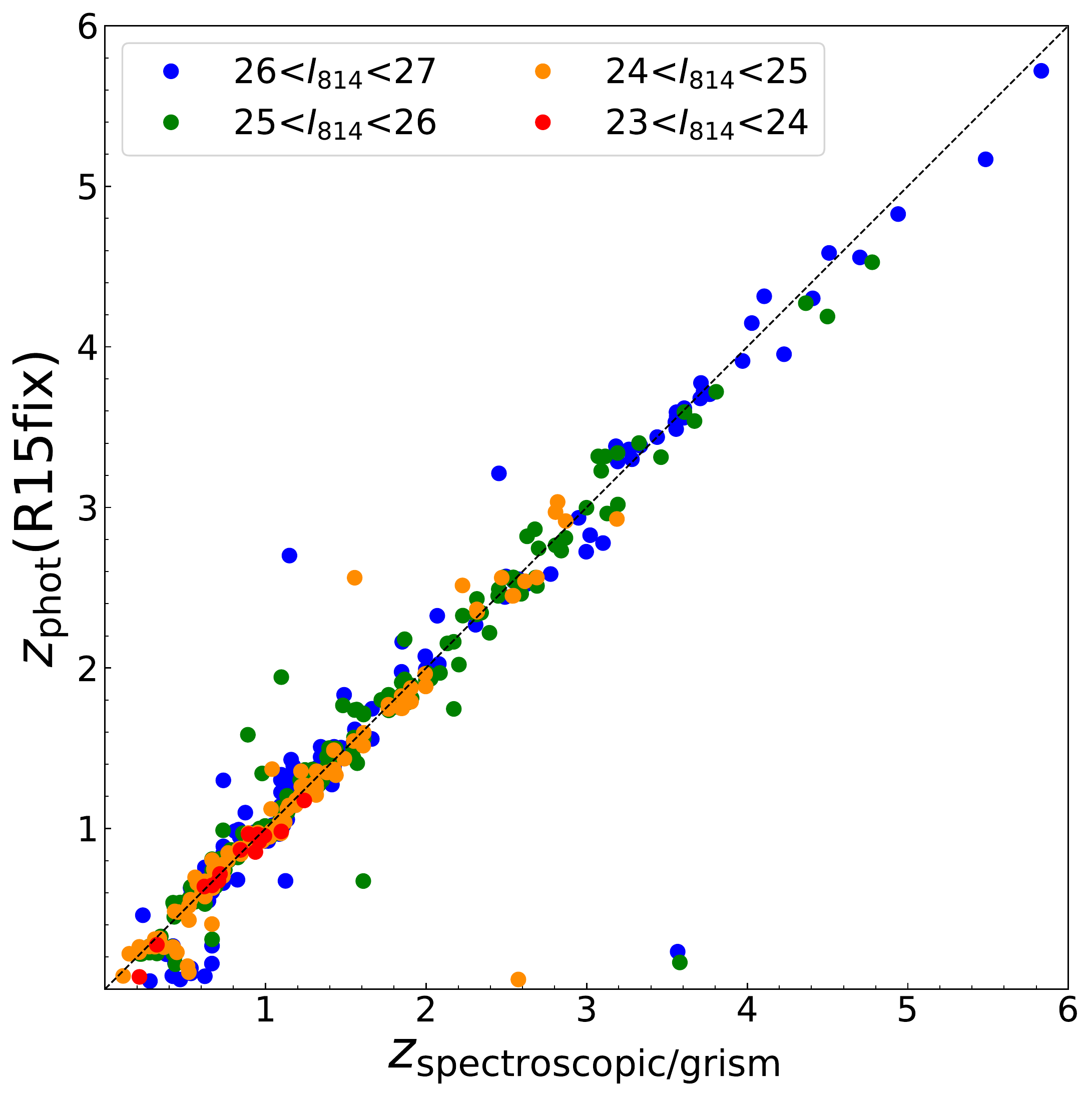}
\caption {Similar to Figure \ref{fig:1} but with purely magnitude selection applied. Different colours correspond to magnitude ranges as indicated.}
\label{fig:2}
\end{figure*}

\begin{figure*}
 \includegraphics[width=0.6\linewidth]{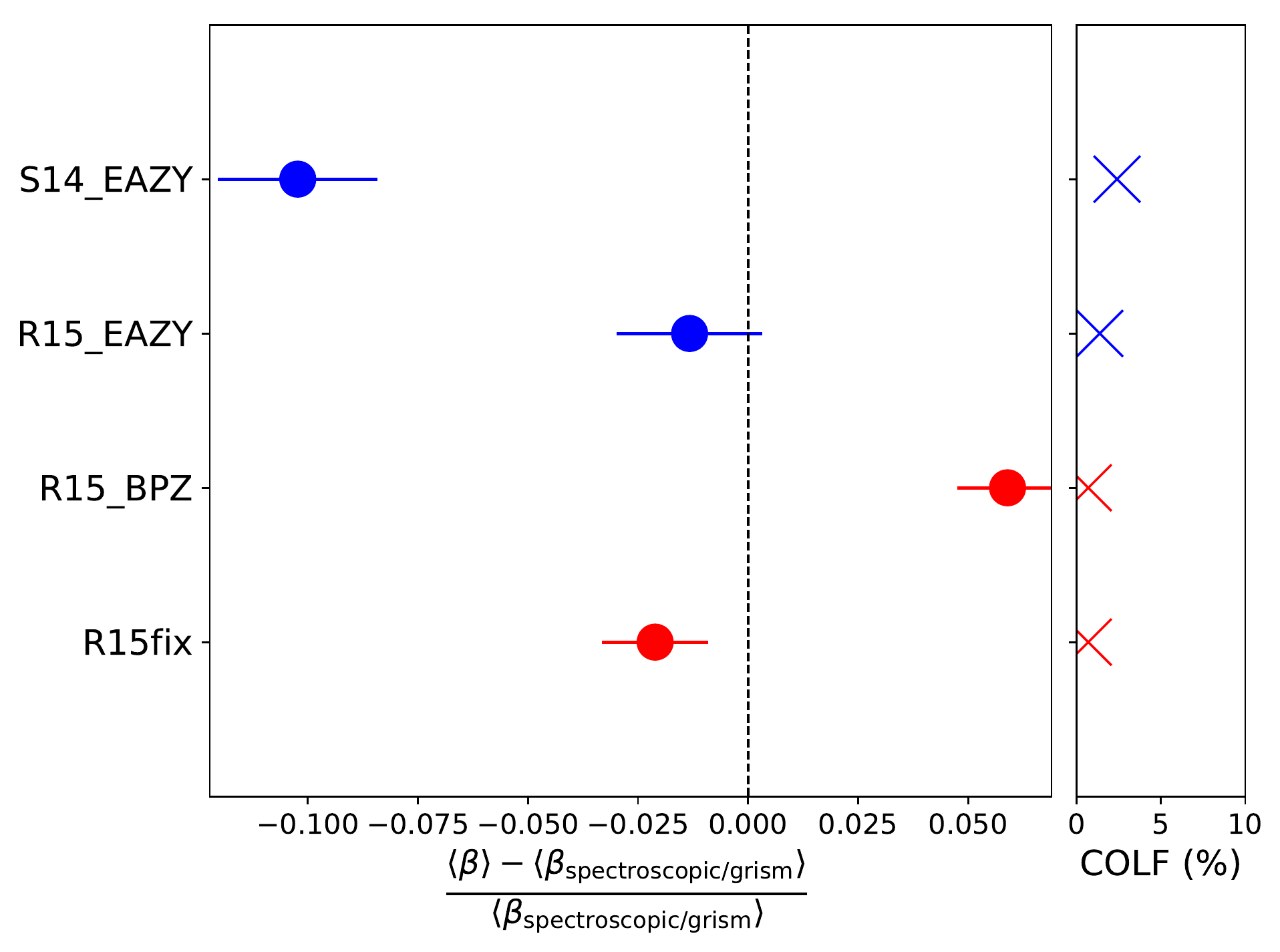}
\caption{The relative bias in the mean geometric lensing efficiency normalised to the $z_{\mathrm{spec/grism}}$ of our purely magnitude-selected sample. The errors are from bootstrapping (see text). The right panel shows the corresponding COLF (see text).
\label{fig:nmad2}}
\end{figure*}

Although the redshift offset in \fixphotz ~is formulated using the colour-magnitude-selected sample, we find that it is still appropriate for the purely magnitude-selected sample. The corresponding plots for the purely magnitude-selected sample are shown in Figures \ref{fig:2} and \ref{fig:nmad2}. 

The overall trend is quiet similar to what is seen in the colour-magnitude-selected sample. Applying the shifts is seen to slightly overcompensate the biased \betas (refer to label \texttt{R15fix} in Figure \ref{fig:nmad2}). In Figure \ref{fig:2} we see that \skeltonphotz ~indeed suffers from outliers that are catastrophically biased low including fairly bright galaxies with \mbox{$24<I_{\mathrm{814}}<25$}. Using deeper data with \eazy ~reduces the catastrophic outliers but also does not completely remove them (refer to label $z_{\mathrm{phot}}\texttt{(R15\_EAZY)}$ in Figure \ref{fig:2}). \fixphotz ~have the least remaining catastrophic outliers, which confirms our choice of using these photo-$z$s as our new calibration sample. Different to the spec/grism-$z$s sample it does not suffer from incompleteness at relevant depths needed for weak lensing studies.

 \begin{figure*}   
 \includegraphics[width=\linewidth]{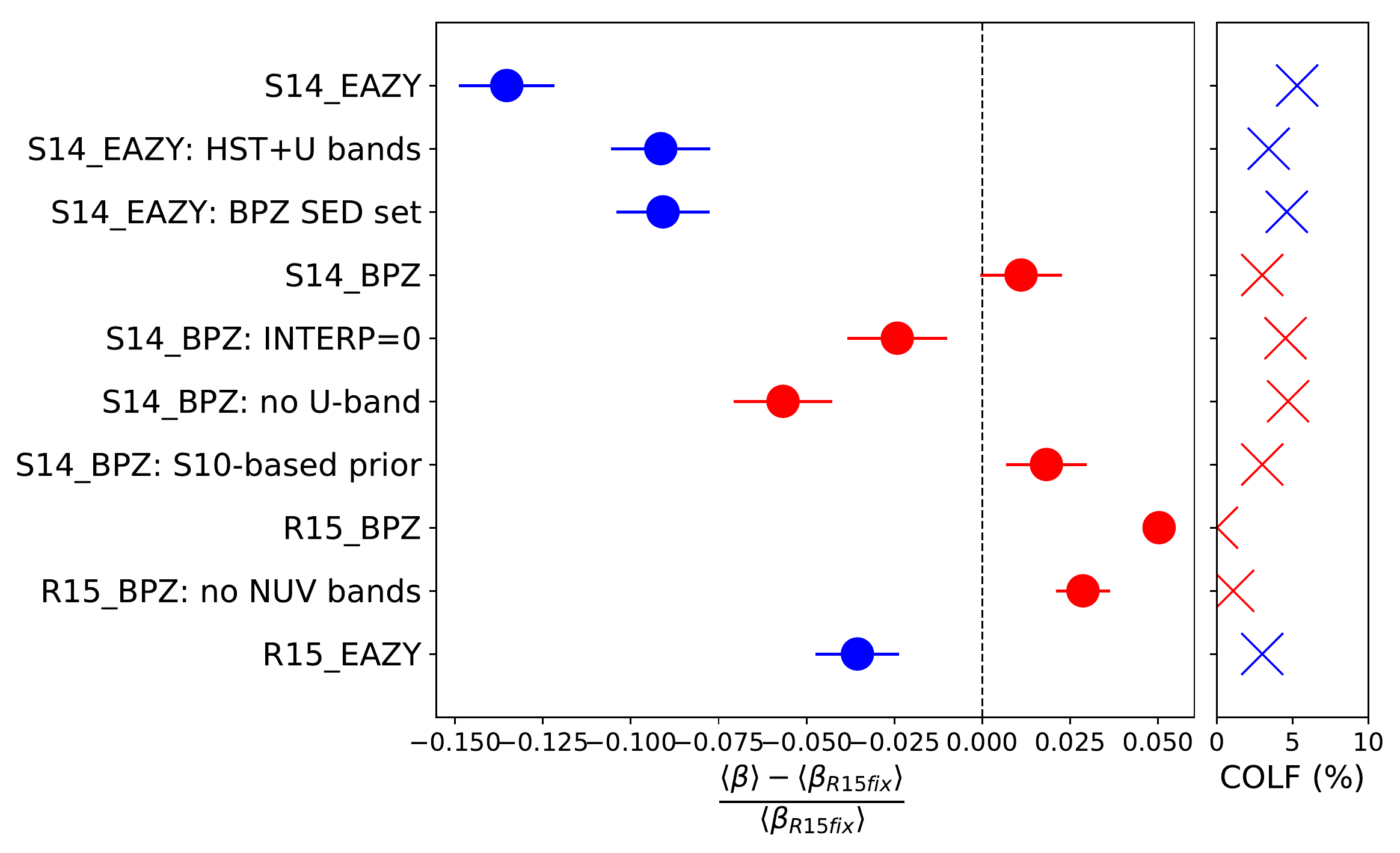}
 \caption{Similar to Figure \ref{fig:vccut} but for a purely magnitude-selected sample.
\label{fig:vfull}}
\end{figure*}

 \begin{figure*}   
 \includegraphics[width=\linewidth]{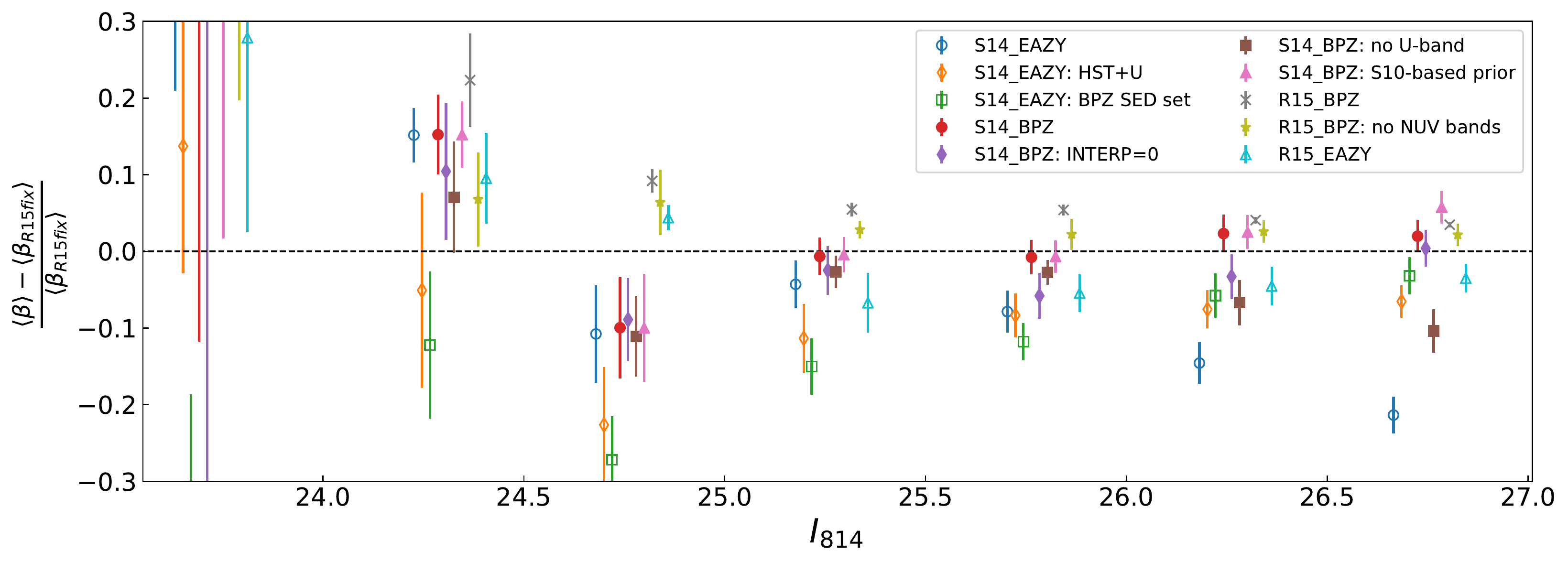}
\caption{Similar to Figure \ref{fig:vccutmagbin} but for a purely magnitude-selected sample.
\label{fig:vfullmagbins}}
\end{figure*}

\begin{figure*}
\includegraphics[width=0.68\columnwidth]{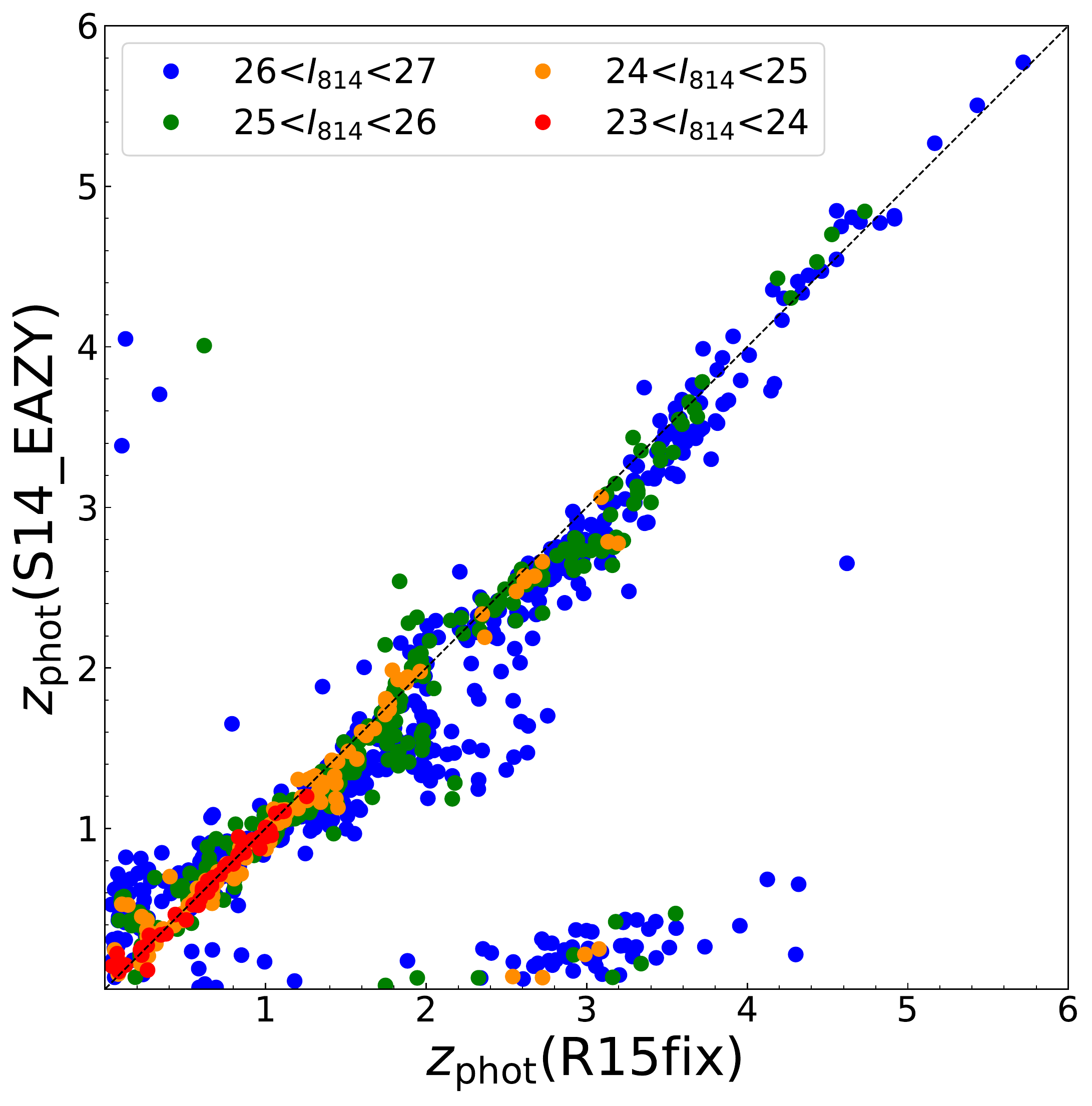}
\includegraphics[width=0.68\columnwidth]{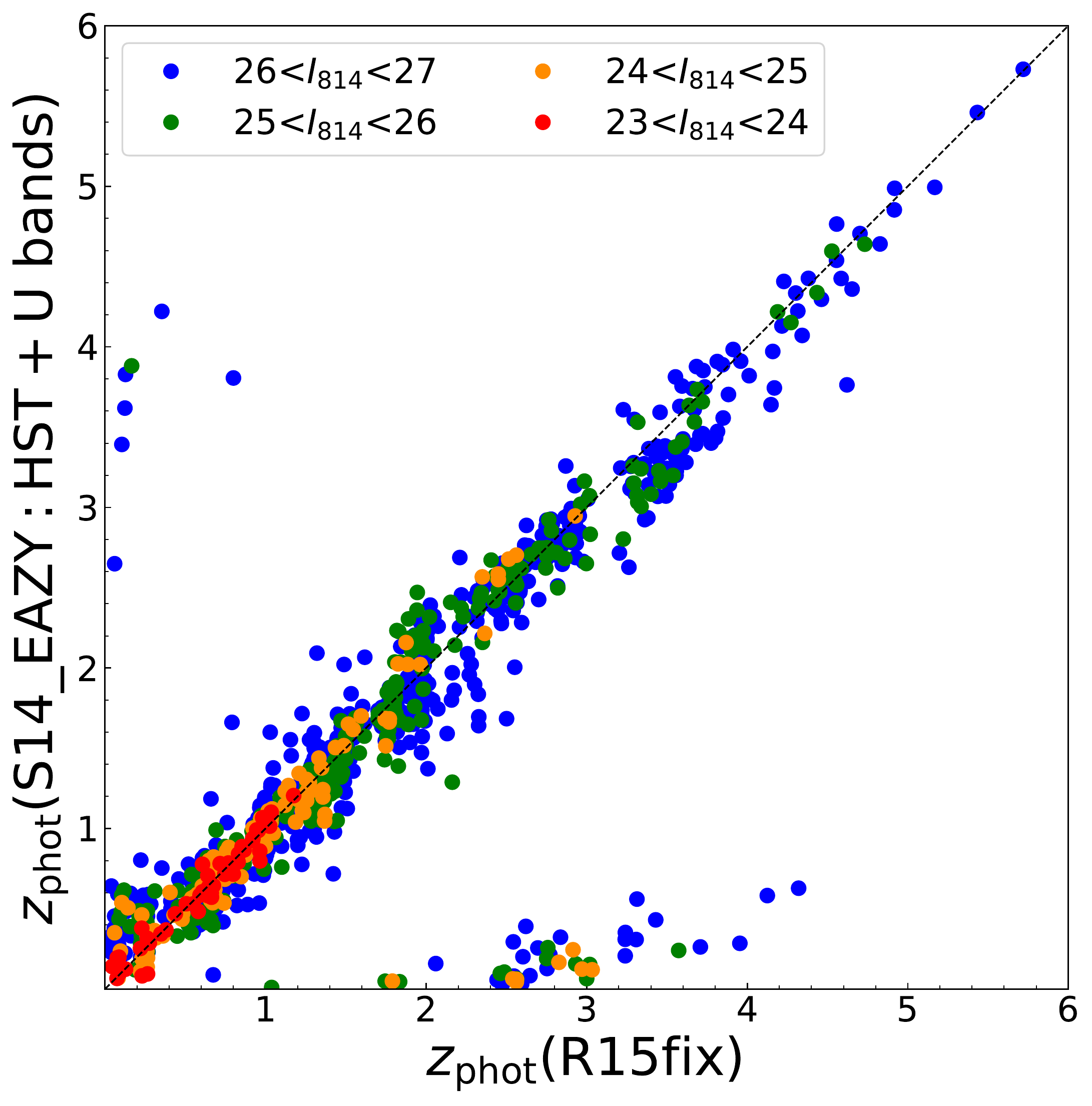}
\includegraphics[width=0.68\columnwidth]{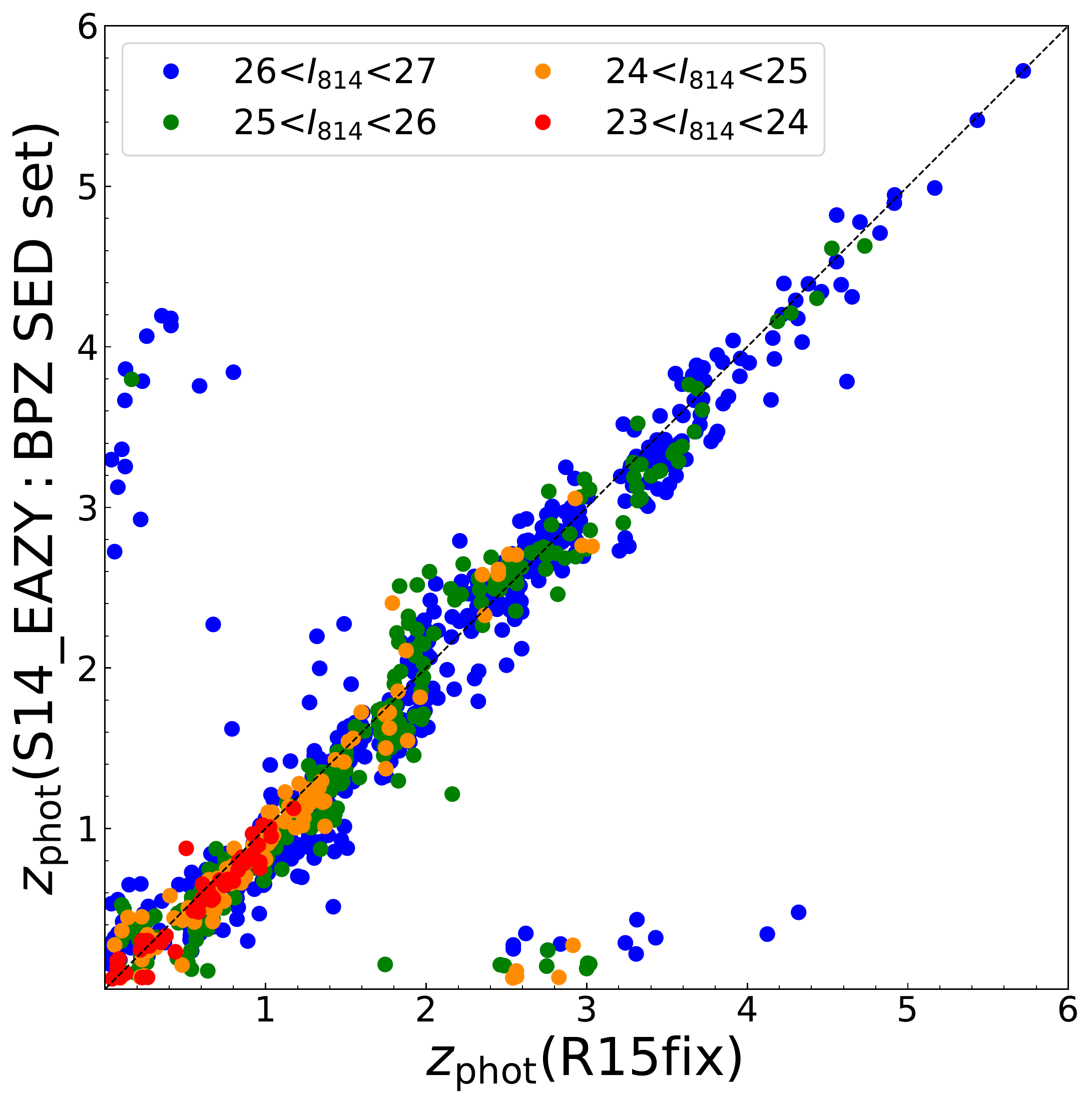}
\includegraphics[width=0.68\columnwidth]{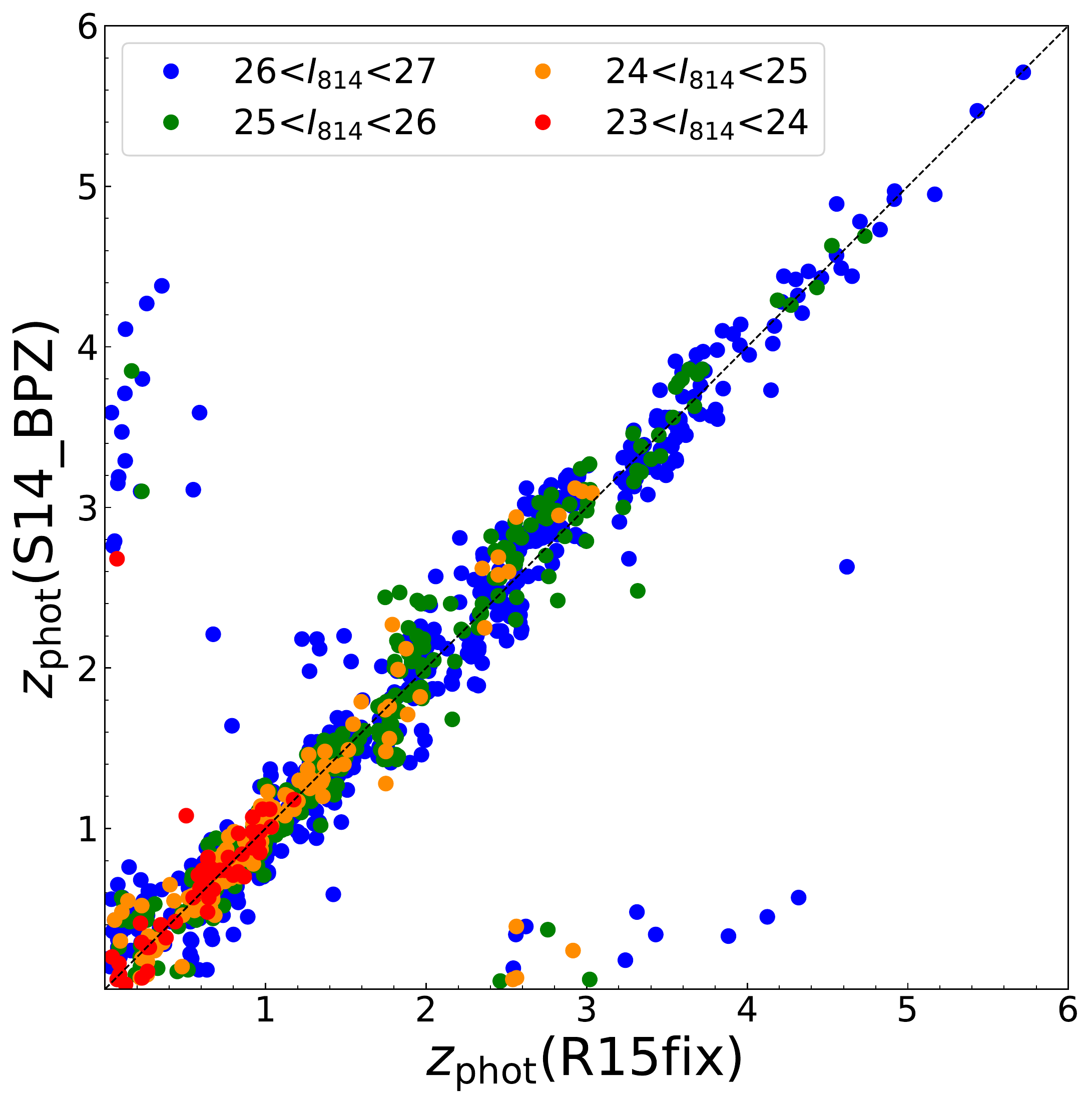}
\includegraphics[width=0.68\columnwidth]{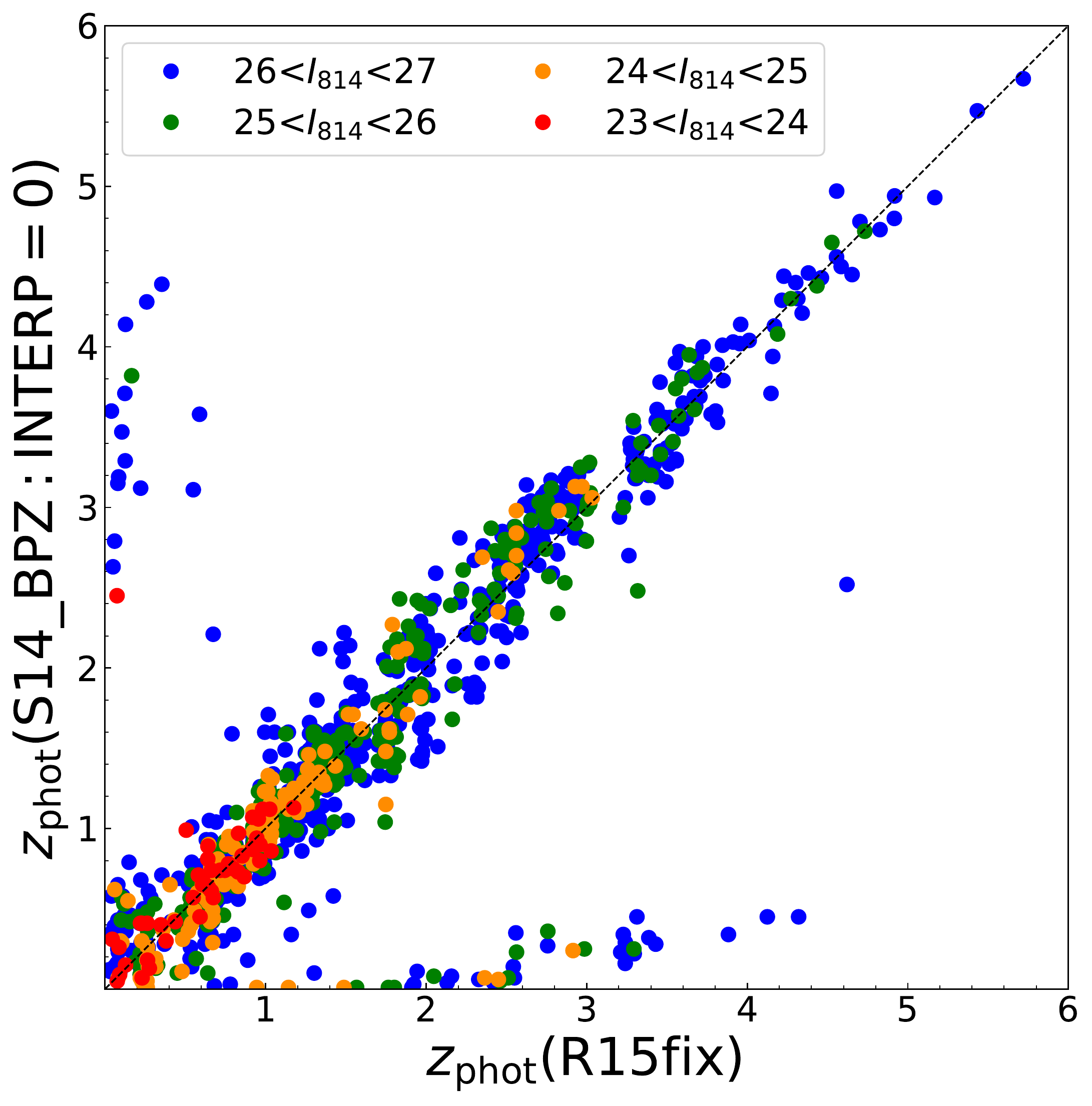}
\includegraphics[width=0.68\columnwidth]{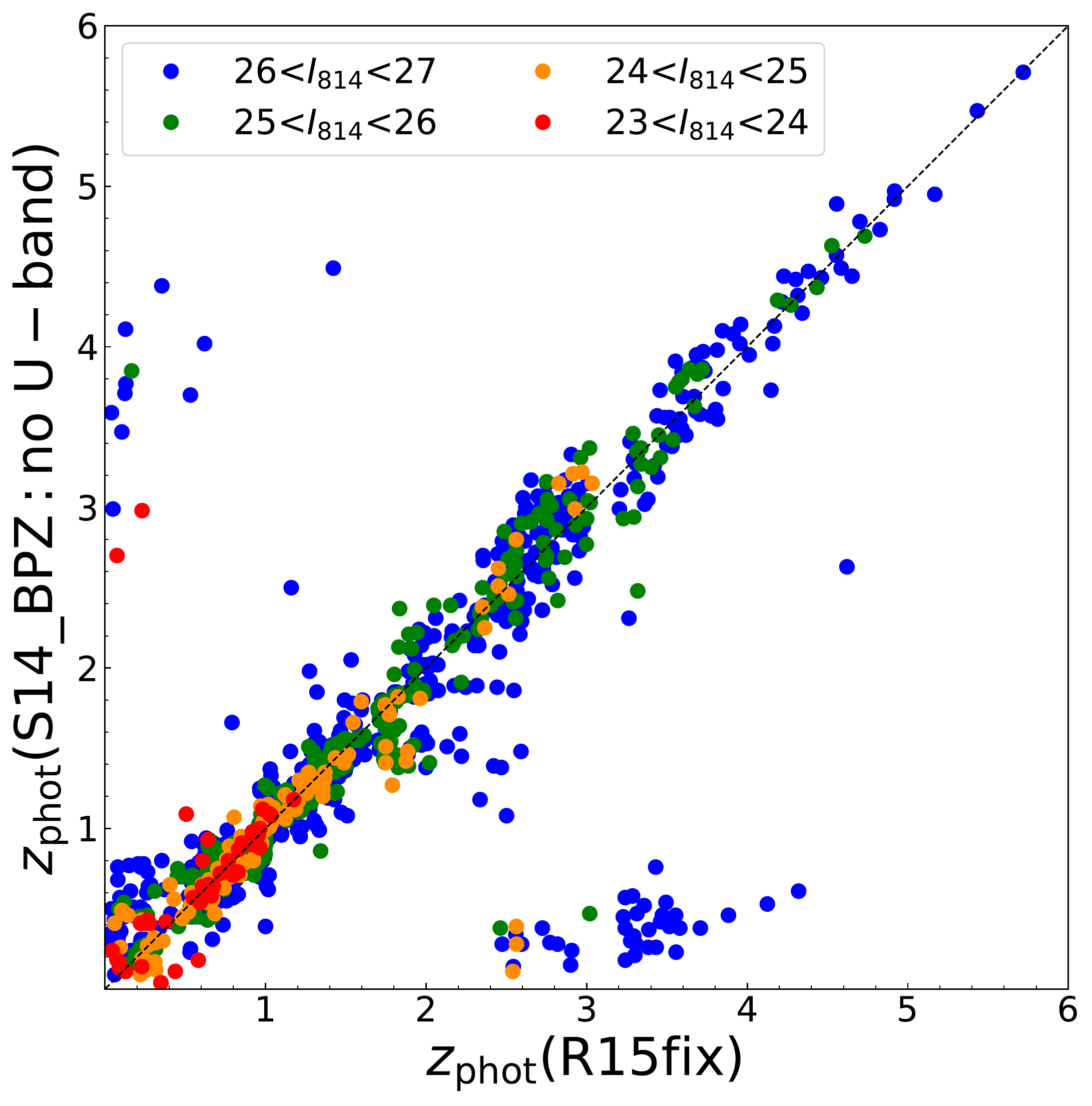}
\includegraphics[width=0.68\columnwidth]{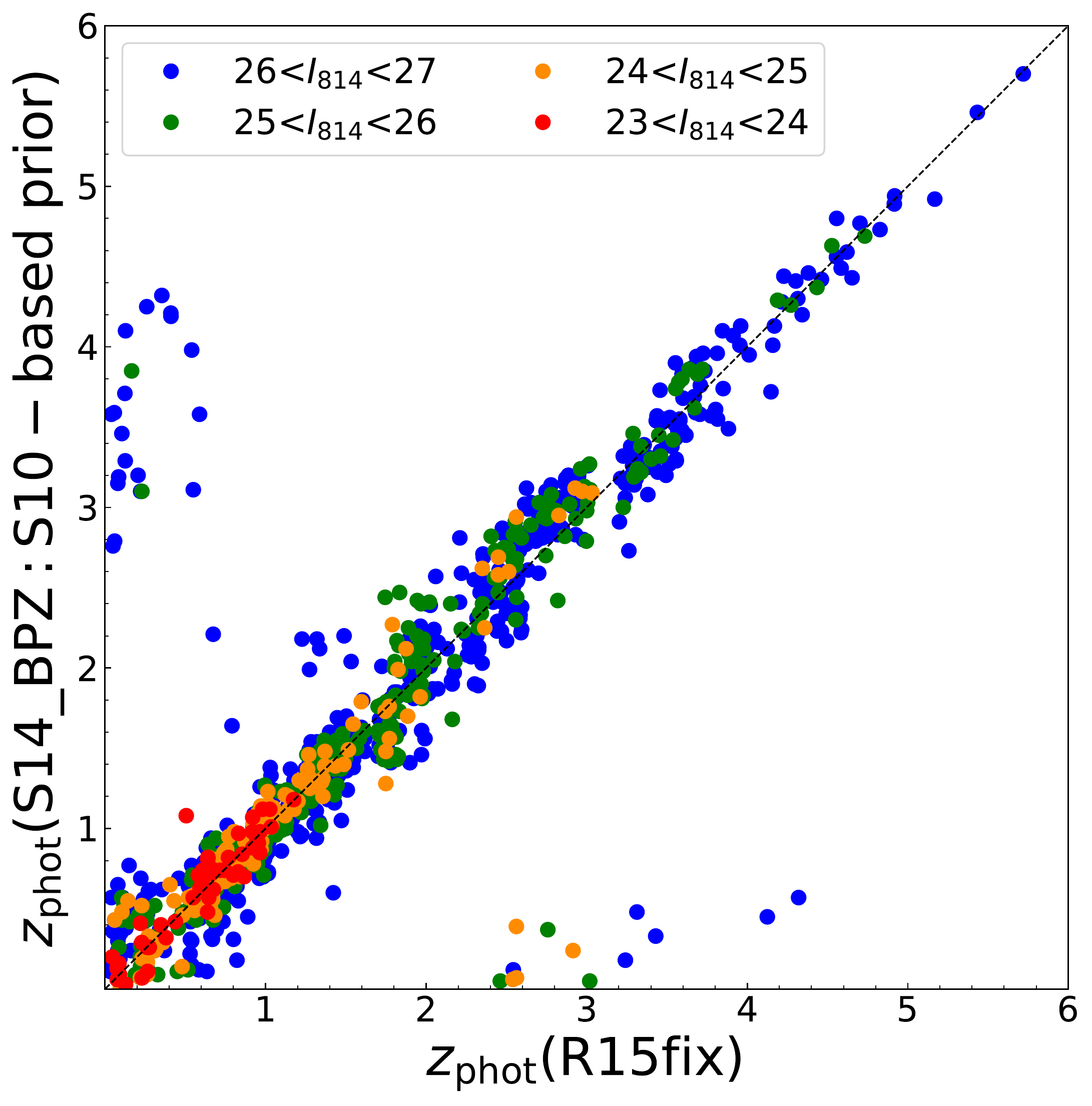}
\includegraphics[width=0.68\columnwidth]{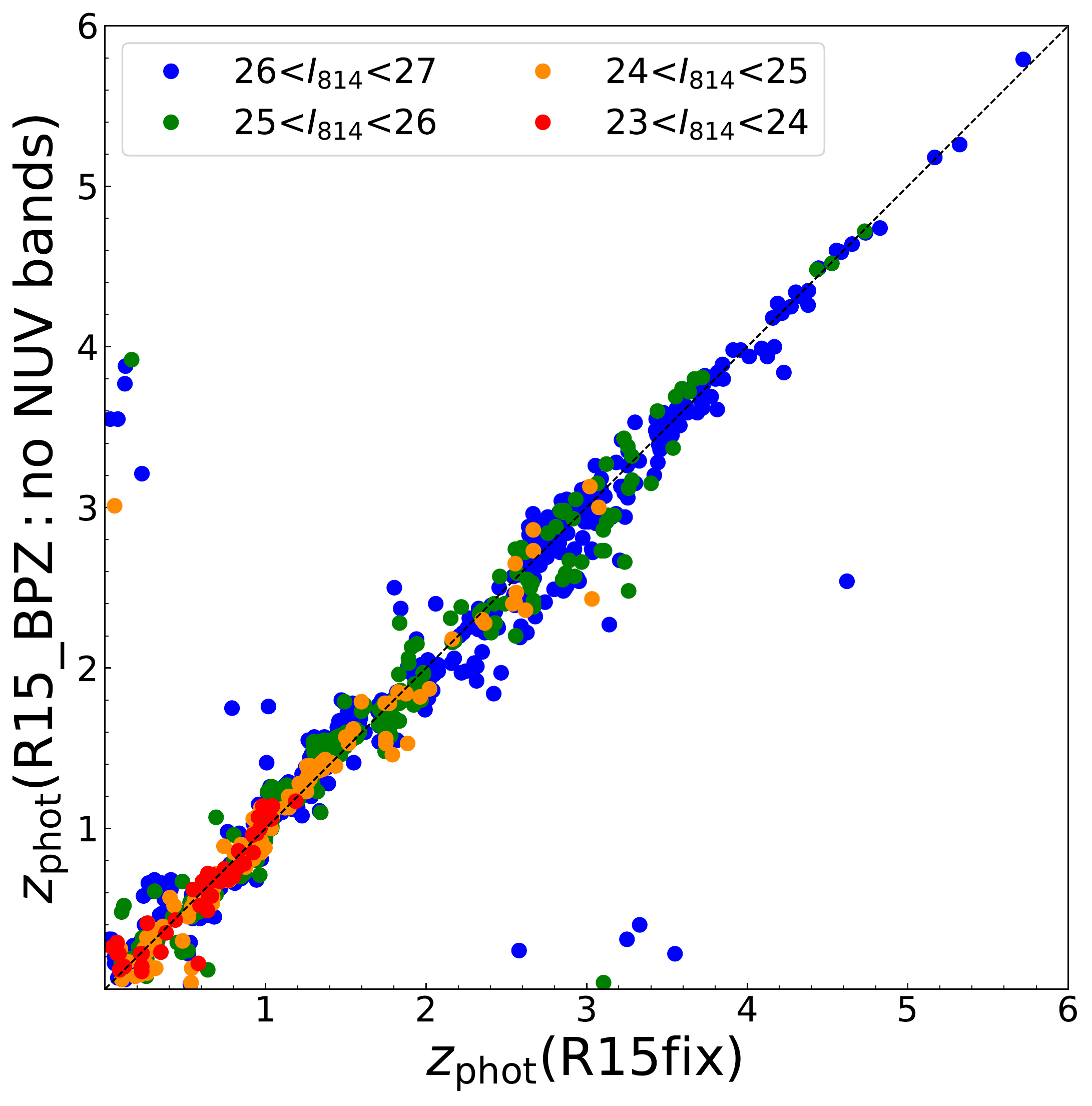}
\includegraphics[width=0.68\columnwidth]{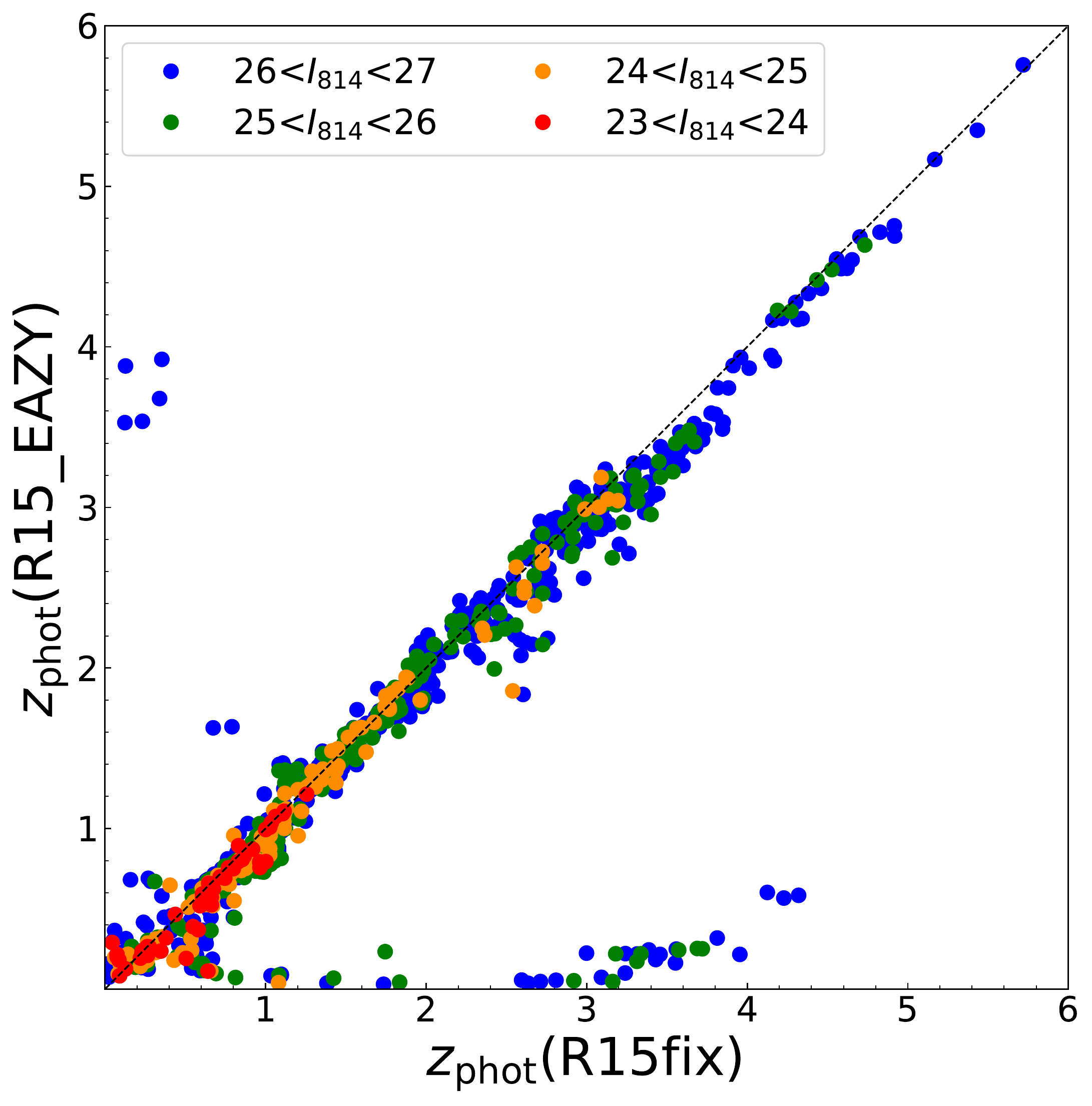}
\caption{Comparison of the purely magnitude-selected sample photo-$z$s to \texttt{R15fix} photo-$z$s with symbols and colours similar to Figure \ref{fig:1}.
\label{fig:fullR15fix1}}
\end{figure*}

The total relative bias and error, COLF, relative bias in magnitude bins, and redshift comparison plots for the purely magnitude-selected sample are shown in Figure \ref{fig:vfull}, \ref{fig:vfullmagbins} and \ref{fig:fullR15fix1}. Overall, the trend that \eazy ~underestimates and \bpz ~tend to slightly overestimate the mean \betas ~is also shown here with biases of $\sim-15\%$ for \skeltonphotz ~and $\sim+5\%$ for \rafelskiphotz. Note that the purely magnitude-selected sample includes significantly fainter galaxies, so we can see here the result of our test on much noisier data. We find that removing ground bands and FIR-bands leads to a bigger improvement for the relative bias compared to the analysis for the colour-magnitude-selected sample. Changing the SEDs set after that did not have any significant impact on the overall bias. However, it turns out that the SEDs used by \eazy ~seem to work better than \bpz's SEDs set for high S/N data, while the opposite is the case for low S/N data (see Figure \ref{fig:vfullmagbins}). This confirms the fact that \bpz's SED are more focused on the high-redshift blue, star-bursting galaxies. We manage to get a relatively unbiased mean \betas ~after switching to \bpz. However, unlike for the colour-magnitude-selected sample this cannot be fully explained by the template interpolation. A possible additional cause could be that the template error function in \eazy, which is an exclusive function in the algorithm, is not suited for the SED templates used by \bpz. Therefore, it cannot function optimally with very low S/N data. 

The impact of $U$-band is more significant in this sample. The difference in the relative bias of the test with $U$-band compared to no $U$-band increases as the $I$-band magnitude increases, especially at the faintest bin where we have the largest number of galaxies. This shows that extending the wavelength range to $U$-band is crucial when studying galaxies that are fainter and more distant in future studies. We see also a small difference in our no NUV bands \bpz ~photo-$z$s and \rafelskiphotz. They mainly differ in the brighter magnitude bins and not in the faint bins. This might be due to the fact that the NUV bands are not as deep as the rest of the \citetalias{Rafelski2015UVUDF:FIELD} data.

Modifying the redshift prior shows to have a very small impact on the photo-$z$ determination for the magnitude-selected sample. 

\bsp	
\label{lastpage}
\end{document}